\documentclass[12pt]{article}
\usepackage{amsmath, amssymb, lamsarrow}
\usepackage{amssymb}
\usepackage{tikz}
\usepackage{hyperref}

\def\mineappendix{
        \setcounter{section}{1}
        \setcounter{subsection}{0}
        \def\thesection{\Alph{section}}
        \def\sectionap{\@startsection  {section}{1}{\z@}
                        {-3.5ex plus-1ex minus-.2ex} {0ex plus.2ex}
                        {\reset@font\Large\bf  Appendix:  \, }
                        }
        }
\makeatother
\def\Proclaim #1. #2\par{\bigbreak\noindent{\sc#1.\enspace}{\it#2}\par}

\newtheorem{lem}{Lemma}[section]
\newtheorem{cor}[lem]{Corollary}
\newtheorem{thm}[lem]{Theorem}
\newtheorem{pro}[lem]{Proposition}
\newtheorem{exa}[lem]{Example}

\newtheorem{rem}[lem]{Remark}

\newcommand{\la}{\lambda}

\author{Xiaobo Liu \thanks{Research was partially supported by NSFC grants 11890662 and 12341105.},
	Chenglang Yang \thanks{Research was partially supported by NSFC grants 12288201, 12401079 and PSFC grants 2023M743717, BX20240407.}}

\date{}

\begin{document}
\title{Action of $W$-type operators on Schur functions and Schur Q-functions}

\maketitle

\begin{abstract}
In this paper, we investigate a series of W-type differential operators, which appear naturally in the symmetry algebras of
KP and BKP hierarchies. In particular, they include all operators in the W-constraints for tau-functions of higher KdV hierarchies which satisfy the string equation. We will give simple uniform formulas for actions of these operators on all ordinary Schur functions and Schur Q-functions.
As applications of such formulas, we will give new simple proofs for Alexandrov's conjecture and Mironov-Morozov's formula, which express the Br\'{e}zin-Gross-Witten and Kontsevich-Witten tau-functions as linear combinations of Q-functions with simple coefficients respectively.
\end{abstract}

%\tableofcontents

\renewcommand{\thefootnote}{\fnsymbol{footnote}} 
\renewcommand{\thefootnote}{\arabic{footnote}}

\section{Introduction}\label{sec:intro}
Ordinary Schur functions and Schur Q-functions give characters for irreducible representations and  projective representations of symmetric groups
respectively (see \cite{Mac}, \cite{Mor}).
They are also polynomial tau-functions for KP and BKP integrable hierarchies respectively (see \cite{DJM}, \cite{Y}).
These functions play fundamental roles in the investigation of super-integrability of matrix models, where many
 partition functions of matrix models are expressed as linear expansions over such character functions (see, for example, \cite{MM20}).
In particular, simple formulas using Q-functions were obtained for Kontsevich-Witten and Br\'{e}zin-Gross-Witten (abbreviated as BGW) tau-functions, which have geometric interpretations as generating functions of certain intersection numbers on moduli spaces of stable curves
 (see \cite{MM20}, \cite{Alex20}, \cite{Alex21}, \cite{LY}, \cite{LY2}).
 An interesting question is whether these character functions and their generalizations can also be used to give simple expansions for
 more general matrix models and Gromov-Witten invariants for compact symplectic manifolds. In fact, it was proposed in \cite{MM21} to use Hall-Littlewood functions, which include Schur functions and Q-functions as special cases, to study generalized Kontsevich matrix models.

Actions of Virasoro operators on Q-functions play an important role in the study of expansions of Kontsevich-Witten and BGW tau-functions (see \cite{LY}, \cite{LY2}). Both of these families of functions are tau-functions of KdV hierarchies which are uniquely determined by the Virasoro constraints.  There are also many models in physics and in geometry where potential functions are tau-functions of  higher KdV (also called Gelfand-Dickey) hierarchies. Important examples include generalized Kontsevich matrix models and generating functions for
intersection numbers on moduli spaces of spin curves (see, for example, \cite{W}, \cite{FSZ}, \cite{MM21}).
For these models, Virasoro constraints no longer uniquely fix the partition functions. Instead, we need a much larger system of constraints, called W-constraints,
to fix the partition functions (see, for example, \cite{AM92}, \cite{FKN}, \cite{Goe}, \cite{Zhou13}). Therefore to investigate the super-integrability
of more general models, it is important to know how W-type operators act on character functions and their generalizations.

The main purpose of this paper is to give concise formulas for the actions of a series of differential operators
$\widetilde{P}^{(k)}_m$ and $\widehat{P}^{(k)}_m$ on Schur functions and Q-functions, respectively.
These operators belong to the algebras $W_{1+\infty}$ or
$W_{1+\infty}^B$, which give symmetry for KP and BKP hierarchies respectively
(see, for example, \cite{FKN} and \cite{Alex21}).
After some modifications, they give Virasoro operators,
 cut-and-join operators (see, for example, \cite{MS}, \cite{Alex11}, \cite{Alex18}, \cite{MMMR}), and all operators in the W-constraints for tau-functions of higher KdV hierarchies which
 satisfy the string equation (see \cite{AM92}, \cite{FKN}, and \cite{Goe}).
We hope that the method used in this paper will also help to study actions of similar operators on
more general Hall-Littlewood functions.
In fact, a key ingredient in the proofs of these formulas is a vertex operator representation
for operators $P^{(k)}_m(\rho)$, which will be given in Theorem \ref{thm:O as BB}. This theorem also holds
for all Hall-Littlewood functions.

Given $\la=(\la_1, \ldots, \la_l) \in \mathbb{Z}^l$, we will denote the associated Schur function and Q-function
by $S_\la(\mathbf{t})$ and $Q_\la(\mathbf{t})$ respectively. These functions are polynomials
in the variables $\mathbf{t}=(t_1,t_2,...)$. In the theory of symmetric functions,
$n t_n$ is the $n$-th power sum function of another set of infinitely many variables (see \cite{Mac}).
Via a standard procedure, all functions $S_\la$ (respectively $Q_\la$) can be represented as linear combinations of
those $S_\mu$ (respectively  $Q_\mu$) with $\mu$ being partitions (see Section \ref{sec:pre} for details).

We introduce a series of differential operators $P^{(k)}_m(\rho)$ with $k \in \mathbb{Z}_{+}, m \in \mathbb{Z}, \rho\in\mathbb{C}$, whose
 generating series $P^{(k)}(z;\rho)=\sum_{m\in\mathbb{Z}}P^{(k)}_m(\rho) \, z^{-m-k}$ are given by
\begin{align}\label{eqn:def O}
P^{(k)}(z;\rho) \,= \,\,:(\partial_z+J(z;\rho))^{k-1} J(z;\rho):,
\end{align}
where "$: \,\,\, :$" is the normal ordering for products of differential operators and
\begin{align}\label{eqn:def:J}
J(z;\rho)
 =\sum_{n\in\mathbb{Z}_{+}} (1-\rho^n)n t_n \, z^{n-1} +\sum_{n\in\mathbb{Z}_{+}} \frac{\partial}{\partial t_n} \, z^{-n-1}.
\end{align}
We will set $\rho=0$ or $\rho=-1$ when considering actions of those operators on Schur functions or Q-functions respectively.
Reference to parameter $\rho$ will be dropped if there is no confusion from the context.

The first main result of this paper gives actions of operators $P^{(k)}_m(-1)$ on Q-functions.
\begin{thm}\label{thm:main2}
When $\rho=-1$,
we set $\widehat{P}^{(k)}_m=P^{(k)}_m(-1)$.
For any $k\in\mathbb{Z}_+, m\in\mathbb{Z}$, and $\la=(\la_1,\dots,\la_l)\in\mathbb{Z}^l$,
the action of $\widehat{P}^{(k)}_m$ on $Q_\la$ is given by
\begin{align}\label{eqn:main2}
\widehat{P}^{(k)}_m \cdot Q_\la
=\sum_{i=1}^l c_{k}^{m}(\la_i) Q_{\la-m\epsilon_i}
+\delta_{m<0} \sum_{a=0}^{-m} (-1)^{m-a} d_k(a)Q_{(\la,a,-m-a)},
\end{align}
where $\la-m\epsilon_i :=(\la_1,\dots,\la_i-m,\dots,\la_l)$, and for any integer $n$,
\begin{align}\label{eqn:d2}
d_k(n) := k! \frac{(-1)^{k-1} }{2^{k}} \sum_{j=0}^{k-1} (-2)^j  \binom{n}{j},
\end{align}
\[c_{k}^m(n) := 2d_k(n-m) -2(-1)^{m}d_k(-n).\]
\end{thm}
In this paper, we set $\binom{n}{k}:= \frac{[n]_k}{k!}$ for any integer $n$ and non-negative integer $k$, where
$$ [n]_k := \prod_{j=0}^{k-1} (n-j) $$
is the falling factorial of length $k$ for integer $n$.  We set $[n]_0:=1$ for all $n$.
In particular, $n$ could be negative here. The constant $\delta_{m<0}$ is defined to be equal to $1$ if $m<0$ and $0$ otherwise.

The $k=1$ case of the above theorem
gives well-known formulas for $\frac{\partial}{\partial t_n} Q_\la$ and $t_n Q_\la$.
%For example, the formula for $\frac{\partial}{\partial t_n} Q_\la$ can be found in Example 11 on page 265  in \cite{Mac}.
%The formula for $t_n Q_\la$ can be found in \cite{B} and \cite{LY}.
When $k=2$, this theorem gives action of Virasoro operators on $Q_\la$ which was obtained
in \cite{ASY}, \cite{LY}, \cite{LY2}, \cite{ASY2} for various special cases (see Example \ref{exm:VirQ} for more details).
The $k=3$ case of this theorem is also very interesting since certain combinations of operators
$\widehat{P}^{(k)}_m$ with $k \leq 3$ for $m=-1$ and $m=-3$ give the cut-and-join operators for BGW and Kontsevich-Witten tau-functions
respectively.
In Section \ref{sec:pf2app}, we will use the formula for $m=-1$  to
 give  a new simple combinatorial proof for Alexandrov's conjecture
which represents BGW tau-function as a linear combination of Q-functions with simple coefficients (see \cite{Alex20}).
This conjecture has been proved independently
 in \cite{Alex21} and \cite{LY2} via different methods.
In Section \ref{sec:pf KW}, we will use the formulas for $m=\pm3$ to give a new proof of Mironov-Morozov's formula, which expresses Kontsevich-Witten tau-function in terms of Q-functions with simple coefficients (see \cite{MM20} and \cite{Alex20}). This proof is simpler than the proof
using Virasoro constraints given in \cite{LY} and also does not need matrix model.

The second main result of this paper gives the action of operators $P^{(k)}_m(0)$ on Schur functions.
\begin{thm}\label{thm:main1}
When $\rho=0$,
we set $\widetilde{P}^{(k)}_m=P^{(k)}_m(0)$.
For any $k \in \mathbb{Z}_{+}, m \in\mathbb{Z}, \la=(\la_1,\dots,\la_l) \in \mathbb{Z}^l$, we have
\begin{align}\label{eqn:main1}
\widetilde{P}^{(k)}_m\cdot S_\la
= & \sum_{i=1}^l k [\la_i-m-i]_{k-1} \, S_{\la-m\epsilon_i}
+\delta_{m,0} [-l]_k \, S_\la  \nonumber \\
& +\delta_{m<0}\sum_{n=1}^{-m} (-1)^{m-n} k[n-l-1]_{k-1} \, S_{(\la,n,1^{-m-n})},
\end{align}
where $(1^n):=(1,\dots,1) \in \mathbb{Z}^n$ for any positive integer $n$
and $1^0=\emptyset$ is the empty partition.
% \begin{align}\label{eqn:def:c1}
% c^{(k)}(a)=k\cdot \prod_{j=1}^{k-1}(a-j+1),
% \end{align}
%\begin{align}\label{eqn:def:d1}
% h_{k}(l,n) := k! \sum_{j=0}^{k-1} (-1)^j \binom{n-1}{k-1-j} \binom{l-1+j}{j}.
%\end{align}
\end{thm}

The $k=1$ case of this theorem corresponds to well-known formulas for $\frac{\partial}{\partial t_n} S_\la$ and $t_n S_\la$. The $k=2$ case of this theorem gives actions of Virasoro operators on $S_\la$, which was obtained
in \cite{LY3}. Another formula with complicated coefficients for the action of Virasoro operators on Schur functions was also given in \cite{AM}.  In the $k=3$ and $m=0$ case, the operators considered here
give the cut-and-join operator for the generating function of Hurwitz numbers (see, for example, \cite{KL} and \cite{Zhou}).
The action of this cut-and-join operator on Schur functions was given in \cite{G}, \cite{GJ}, \cite{FW}, and \cite{Zhou}.
 Formulas for $\widetilde{P}^{(3)}_m S_\la$ with $m=-1$ or $m=-2$ were used in \cite{MMMR} and \cite{WZZZ} to study Hermitian matrix models.

Note that the operators $\widehat{P}^{(k)}_m$ and $\widetilde{P}^{(k)}_m$ contain infinitely many terms, each of them is a product of $k$ operators
like $t_n$ or $\frac{\partial}{\partial t_n}$. One might expect that $\widehat{P}^{(k)}_m Q_\la$ and $\widetilde{P}^{(k)}_m S_\la$
would be very complicated in general. It is surprising that the formulas given in the above two theorems
are rather simple and have uniform expressions for all $k$ and $m$.
In fact,  the right hand sides of formulas \eqref{eqn:main2} and \eqref{eqn:main1}
 contain at most $l(\la)+|m|$ terms where $l(\la)$ is the number of components in $\la$.
 We can also see from these formulas that $Q_\la$ and $S_\la$ are eigenfunctions of $\widehat{P}^{(k)}_0$ and $\widetilde{P}^{(k)}_0$, respectively, for all $k$.

This paper is organized as follows. In section \ref{sec:pre}, we review the vertex operator realization of Hall-Littlewood functions, which include Schur functions and Q-functions as special cases.
In Section \ref{sec:pf0}, we give a vertex operator representation for operators $P^{(k)}_m$.
Proofs for Theorem \ref{thm:main2} and Theorem \ref{thm:main1} (and some applications of these theorems) will be given in Section \ref{sec:pf2} and Section \ref{sec:pf1} respectively.

\section{Preliminaries}\label{sec:pre}

There are many definitions of Schur functions and Q-functions (see, for example, \cite{Mac} and \cite{H}).
To give a unified treatment for these functions, we will consider them
as special cases of the Hall-Littlewood functions.
In this section, we review the vertex operator realization for Hall-Littlewood functions  and some of their basic properties.
We will also describe  relations between operators $P^{(k)}_m$ and vertex operators.

In this paper, Hall-Littlewood functions $H_\la(\mathbf{t};\rho)$ are considered as polynomials
 in the variables $\mathbf{t}=(t_1,t_2,...)$ labeled by $\la\in\mathbb{Z}^l$ and parameter $\rho\in\mathbb{C}$.
The vertex operator realization for these functions was introduced in \cite{J1}
(See also \cite{Z}, \cite{H}, \cite{J2}, \cite{Mac} for vertex operator realizations of Schur functions and Q-functions).
We mainly follow the conventions in \cite{JL} for the definition of vertex operators.

Let $B_n, n\in\mathbb{Z}$, be differential operators acting on the ring of polynomials $\mathbb{C}[\mathbf{t}]$, whose generating series
 $B(z):=\sum_{n\in\mathbb{Z}}B_n z^n$ is defined by
\begin{align}\label{eqn:def B}
B(z):=\exp\Bigg(\sum_{n\in\mathbb{Z}_{+}}(1-\rho^n)t_n z^n\Bigg) \exp\Bigg(-\sum_{n\in\mathbb{Z}_{+}} \frac{1}{n}\frac{\partial}{\partial t_n} z^{-n}\Bigg).
\end{align}
Note that $B_n$ depends on the choice of a parameter $\rho \in \mathbb{C}$.
For any fixed $\rho$, $B(z)$ is a distribution in indeterminate $z$, i.e., $B(z) \in V[\![z,z^{-1}]\!]$, where $V$ is the linear space of differential operators on the polynomial ring $\mathbb{C}[\mathbf{t}]$ (see Chapter 2 in \cite{K} for an introduction for distributions).

For any $\la=(\la_1,...,\la_l)\in\mathbb{Z}^l$, the Hall-Littlewood function associated with $\la$ is defined by
\begin{align}\label{eqn:def HL}
H_\la(\mathbf{t};\rho):=B_{\la_1}\cdots B_{\la_l}\cdot 1 \in\mathbb{C}[\mathbf{t}].
\end{align}
Assign $\deg t_n=n$ and $\deg \frac{\partial}{\partial t_n}=-n$. Then
 $B_n$ is a differential operator of degree $n$ and $H_\la(\mathbf{t}; \rho)$ is a homogeneous polynomial of degree
 $|\la|:=\sum_{i=1}^l \la_i$.
 In particular, $B_n \cdot 1 = \delta_{n,0}$ if $n \leq 0$.
 Hence $H_{(\la,0)}(\mathbf{t};\rho)=H_\la(\mathbf{t};\rho)$  for all $\la \in \mathbb{Z}^l$ and
 $H_\la(\mathbf{t};\rho)=0$ if $\la_l<0$.

Operators $B_{n}$ satisfy the following relation (see Proposition 2.1 in \cite{JL})
\begin{align}\label{eqn:commutation relation for HL:BB}
B_{m}B_{n}-\rho B_{n}B_{m}=\rho B_{m+1}B_{n-1}-B_{n-1}B_{m+1}.
\end{align}
Repeatedly applying this relation, we can represent $H_{\la}$ for
 every $\la \in \mathbb{Z}^l$ as a linear combination of functions $H_\mu$ with $\mu$ being partitions, i.e.
 components of $\mu$ are positive and arranged in an non-increasing order
(see, for example, Lemma 1 in \cite{Mor64} and Example 2 in Section III.2 of \cite{Mac} ).
When considering actions of differential operators on such functions, it is very convenient to allow
$\la$ to be not necessarily a partition.
This extension has been widely used in the literature (see, for example, \cite{L} and \cite{Mor}).

Schur functions and Q-functions are Hall-Littlewood functions with $\rho=0$ and $\rho=-1$ respectively, i.e.
\[ S_\la (\textbf{t}) = H_\la(\textbf{t}; 0), \hspace{20pt} Q_\la (\textbf{t}) = H_\la(\textbf{t}; -1).\]
In these cases, equation \eqref{eqn:commutation relation for HL:BB} is much simpler.
In fact,  we have, for all $m,n \in \mathbb{Z}$,
\begin{equation} \label{eqn:BB-r0}
B_{m}B_{n} = -B_{n-1}B_{m+1}
\end{equation}
for Schur functions (where $\rho=0$),
and
\begin{align} \label{eqn:BB=-BB+delta}
B_m B_n = - B_{n}B_{m}+2(-1)^m \delta_{m,-n}
\end{align}
for Q-functions (where $\rho=-1$,
 see Example 8 on page 263 in \cite{Mac}).
Hence it  is  much easier to
represent $S_\la$ (respectively $Q_\la$) as a linear combination of $S_\mu$
( respectively $Q_{\mu}$)  with $\mu$ being partitions.

Since multiplication operators $t_n$ and derivation operators $\frac{\partial}{\partial t_n}$ do not commute,
we can not simply add the two exponents on the right hand side of equation \eqref{eqn:def B} to form a single exponential function.
To avoid such problems, it is convenient to introduce the normal ordering $``: \,\, :"$ for products of differential operators,
which always shift derivation operators $\frac{\partial}{\partial t_n}$ to the right of multiplication operators $t_n$.
For example, for all $m$ and $n$,
\begin{align}\label{eqn:def::}
:t_n\frac{\partial}{\partial t_m}:
\,\, = \,\, :\frac{\partial}{\partial t_m}t_n:
\,\, = \,\, t_n\frac{\partial}{\partial t_m},\ \ \ \ \text{and\ \ }
:e^{t_n+\partial/\partial t_n}: \,\, = \,\, e^{t_n}e^{\partial/\partial t_n}.
\end{align}
This normal ordering has been used in Chapter 5.2 in \cite{DJM}.
Equation \eqref{eqn:def::} is exactly the one given in Example 5.4 in \cite{DJM}.
As a result, we can rewrite equation \eqref{eqn:def B} as
\begin{align}
B(z) = \,\, :e^{\phi(z)}:,
\end{align}
where $\phi(z)$ is defined by
\begin{align}
\phi(z) := \sum_{n\in\mathbb{Z}_+} (1-\rho^n)t_n \, z^n-\sum_{n\in\mathbb{Z}_+} \frac{1}{n}\frac{\partial}{\partial t_n} \, z^{-n}.
\end{align}
Recall that $J(z)=J(z; \rho)$ is defined by equation \eqref{eqn:def:J}.
We can use the fact
\begin{align}\label{eqn:d phi=J}
\partial_z \phi(z) = J(z)
\end{align}
 to obtain
the following representation for the generating series $P^{(k)}(z) = P^{(k)}(z;\rho)$ defined by equation \eqref{eqn:def O}:
\begin{lem}\label{lem:O=:phi dphi:}
For all integers $k \geq 1$,
\begin{align*}
P^{(k)}(z) \,\,=\,\, :e^{-\phi(z)}\partial^k_{z}e^{\phi(z)}:.
\end{align*}
\end{lem}
{\bf Proof:}
Since normal ordering of operators is commutative, this lemma is equivalent to
\begin{equation}  \label{eqn:PhiP}
:\partial^{k}_{z}e^{\phi(z)}:
\,\,=\,\, : e^{\phi(z)} P^{(k)}(z) :
\end{equation}
for all $k \geq 1$. We will prove this formula by induction on $k$.

By equation \eqref{eqn:d phi=J}, we have
\[: \partial_z e^{\phi(z)}: \,\,=\,\, : e^{\phi(z)} J(z): \,\,=\,\, : e^{\phi(z)} P^{(1)}(z):.\]
So equation \eqref{eqn:PhiP} holds  for $k=1$.

For any integer $n>1$, assume equation \eqref{eqn:PhiP} holds  for $k=n-1$. Then
\begin{align*}
:\partial^n_{z}e^{\phi(z)}:
\,\,=\,\, &: \partial_z \big( e^{\phi(z)} P^{(n-1)}(z) \big):\\
\,\,=\,\, &:e^{\phi(z)} \big(J(z) P^{(n-1)}(z) + \partial_z P^{(n-1)}(z) \big) :\\
\,\,=\,\, &: e^{\phi(z)} \big(\partial_z +J(z)\big)P^{(n-1)}(z): \\
\,\,=\,\, &: e^{\phi(z)} P^{(n)}(z):.
\end{align*}
So equation \eqref{eqn:PhiP} holds  for $k=n$. The lemma is thus proved.
$\Box$

Define another sequence of differential operators $B^*_n, n \in \mathbb{Z},$
whose generating series $B^*(z)=\sum_{n\in\mathbb{Z}}B^*_nz^{-n}$ is given by
\begin{align}\label{eqn:def B*}
B^*(z)
\,\,=\,\, :e^{-\phi(z)}:
\,\,=\,\, \exp\Bigg(-\sum_{n\in\mathbb{Z}_{+}}(1-\rho^n)t_n z^n\Bigg) \exp\Bigg(\sum_{n\in\mathbb{Z}_{+}} \frac{1}{n}\frac{\partial}{\partial t_n} z^{-n}\Bigg).
\end{align}
Then Lemma \ref{lem:O=:phi dphi:}  can be written as
\begin{align}\label{eqn:O=B*dB}
P^{(k)}(z) \,\,=\,\, :B^*(z)\partial^k_{z}B(z):
\end{align}
for $k\geq1$. For convenience, we also define
\[ P^{(0)}(z) \,\,=\,\, :B^*(z)B(z): = 1 \]
to make equation \eqref{eqn:O=B*dB} true for all $k \geq 0$.

$B(z)$ and $B^*(z)$ are called vertex operators dual to each other.
Operators $B^*_a$ and $B_b$ satisfy the following relation
\begin{align}\label{eqn:commutation relation for HL:B*B}
B_aB_b^*-\rho B^*_bB_a
= \rho B_{a-1}B^*_{b-1} - B^*_{b-1}B_{a-1} +(1-\rho)^2 \delta_{a,b}
\end{align}
for all $a, b \in \mathbb{Z}$ (see Proposition 2.1 in \cite{JL}).
Moreover, the degree of operator $B^*_n$ is $-n$ for all $n \in \mathbb{Z}$ and
$B^*_n \cdot 1 = \delta_{n,0}$ for $n \geq 0$.

In the special case where $\rho=-1$, $B^*_n$ and $B_n$ have a  very simple relation. We first observe that by definition,
 $Q_\la(\mathbf{t})$ does not depend on variables $t_{2k}$ for all $k \geq 1$.
Hence we can write $Q_\la(\mathbf{t})$ as $Q_\la(\mathbf{\hat{t}})$ where
$\mathbf{\hat{t}}:=(t_1, t_3, t_5, \cdots)$ containing only variables $t_{2k-1}$ for all $k \geq 1$.
When considering actions on the polynomial ring $\mathbb{C}[\mathbf{\hat{t}}]$, operators $\phi(z)$ and $J(z)$ with $\rho=-1$
are equivalent to operators
\begin{align}
\hat{\phi}(z) := \sum_{n\in\mathbb{Z}_+} 2 t_{2n-1} \, z^{2n-1}
                    -\sum_{n\in\mathbb{Z}_+} \frac{1}{2n-1}\frac{\partial}{\partial t_{2n-1}} \, z^{-2n+1},
\end{align}
and
\begin{align} \label{eqn:Jhat}
\hat{J}(z) := \sum_{n\in\mathbb{Z}_+} 2(2n-1)  t_{2n-1} \, z^{2n-2}
                    + \sum_{n\in\mathbb{Z}_+} \frac{\partial}{\partial t_{2n-1}} \, z^{-2n}
\end{align}
respectively. Accordingly, in this case, we can also replace $\widehat{P}^{(k)}(z)=\sum_{m \in \mathbb{Z}} \widehat{P}^{(k)}_m z^{-k-m}$ by
$\widehat{M}^{(k)}(z)=\sum_{m \in \mathbb{Z}} \widehat{M}^{(k)}_m z^{-k-m}$ where
$\widehat{M}^{(k)}(z) = \,\, :(\partial_z +\hat{J}(z))^{k-1} \cdot \hat{J}(z) :$.
Operators $B(z)=\sum_{n \in \mathbb{Z}} B_n z^n$ and $B^*(z)=\sum_{n \in \mathbb{Z}} B_n^* z^{-n}$ can be replaced by
$\hat{B}(z)=\sum_{n \in \mathbb{Z}} \hat{B}_n z^n =\,\, :e^{\hat{\phi}(z)} : \,\, $ and
$\hat{B}^*(z)=\sum_{n \in \mathbb{Z}} \hat{B}_n^* z^{-n}= \,\, :e^{-\hat{\phi}(z)} :$
respectively.
Since $-\hat{\phi}(z)=\hat{\phi}(-z)$, we have $\hat{B}^*(z) \,\, = \hat{B}(-z)$. This is equivalent to
\begin{align}\label{eqn:B*=B rho=-1}
\hat{B}^*_{n}=(-1)^n \hat{B}_{-n}
\end{align}
for all $n \in \mathbb{Z}$.

\section{Representing $P^{(k)}_m$ in terms of vertex operators}\label{sec:pf0}

In this section, we prove a formula which expresses $P^{(k)}_m$ in terms of vertex operators $B_n$
and $B_n^*$ without using normal ordering.

The relation between normal ordering and ordinary product of  $B^*(z)$ and $B(w)$
is given by the following formula
\begin{align*}
:B^*(z)B(w): =&  B^*(z)B(w) \cdot\frac{z-w}{z-\rho w},
\end{align*}
where the function $\frac{z-w}{z-\rho w}$ should be expanded as a formal power series in $\frac{w}{z}$
in the region $|z|>|w|$
(see Proposition 2.9 in \cite{J1}).
Comparing coefficients of $z^{-a}w^b$  on both sides of the above equation, we have
\begin{align}\label{eqn::B*B:=B*B-B*B}
:B^*_a B_b: =
\sum_{n=0}^{\infty} \rho^n\big( B_{a-n}^*B_{b-n}
- B_{a-n-1}^*B_{b-n-1}\big)
\end{align}
for all $a,b\in\mathbb{Z}$.

\begin{rem}
The infinite sum on the right hand side of equation \eqref{eqn::B*B:=B*B-B*B} is well-defined as an operator on the polynomial ring $\mathbb{C}[\mathbf{t}]$, since for any $f\in\mathbb{C}[\mathbf{t}]$,
\begin{align*}
B_{b-n}\cdot f=
B_{b-n-1}\cdot f=0
\end{align*}
if $b-n$ is smaller than the degree of $f$.
\end{rem}

\begin{lem}\label{cor::BB:-rho:BB:}
For any  $a,b\in\mathbb{Z}$,
\begin{align*}
:B^*_a B_b: -\rho:B^*_{a-1}B_{b-1}: \,\,=
B_{a}^*B_{b} -B_{a-1}^*B_{b-1}.
\end{align*}
\end{lem}
{\bf Proof:}
By equation \eqref{eqn::B*B:=B*B-B*B},
\begin{align*}
\rho \cdot :B^*_{a-1}B_{b-1}:
=\sum_{n=0}^{\infty} \rho^{n+1}\big( B_{a-n-1}^*B_{b-n-1}
- B_{a-n-2}^*B_{b-n-2}\big).
\end{align*}
Replacing $n$ by $n-1$ on the right hand side of this equation, then subtracting
 this equation from equation \eqref{eqn::B*B:=B*B-B*B}, we obtain the desired formula.
$\Box$

As a result, we obtain the following formula, which expresses operators $P^{(k)}_m$ as vertex operators.
\begin{thm}\label{thm:O as BB}
For any $k\in\mathbb{Z}_{\geq0}, m\in\mathbb{Z}$ and $\rho\neq1$, we have
\begin{align}\label{eqn:O as BB}
P^{(k)}_m
= \sum_{a,b\in\mathbb{Z}\atop a-b=m}  h_{k}(b;\rho) \big(B_{a}^*B_{b} -B_{a-1}^*B_{b-1}\big)
+\delta_{m,0}\frac{\rho^k k!}{(1-\rho)^k},
\end{align}
where
\begin{align} \label{def:h_k}
h_{k}(b;\rho) := k! \sum_{j=0}^{k-1} \frac{\rho^j }{(1-\rho)^{j+1}}\binom{b}{k-j}.
\end{align}
\end{thm}
{\bf Proof:}
Equation \eqref{eqn:O as BB} holds for $k=0$ since $P^{(0)}_m=\delta_{m,0}$ by definition.
Since $\rho \neq 1$, an induction on $k$ shows that
this theorem is equivalent to the following recursion relation
\begin{align}\label{eqn:(1-rho)O-rho O=...}
(1-\rho)P^{(k)}_m-\rho kP^{(k-1)}_m
=&\sum_{a,b\in\mathbb{Z}\atop a-b=m} [b]_k \big(B_{a}^*B_{b} -B_{a-1}^*B_{b-1}\big)
\end{align}
for all $k\geq1$.

Taking coefficients of $z^{-k-m}$  on both sides of equation \eqref{eqn:O=B*dB}, we obtain
\begin{align*}
P^{(k)}_m
=\sum_{a,b\in\mathbb{Z}\atop a-b=m} [b]_k :B^*_aB_b:.
\end{align*}
Thus, the left hand side of equation \eqref{eqn:(1-rho)O-rho O=...} is equal to
\begin{align}
(1-\rho)P^{(k)}_m -\rho kP^{(k-1)}_m
=&(1-\rho)\sum_{a,b\in\mathbb{Z}\atop a-b=m} [b]_k :B^*_aB_b:
-\rho\sum_{a,b\in\mathbb{Z}\atop a-b=m} k [b]_{k-1} :B^*_aB_b: \nonumber \\
=&\sum_{a,b\in\mathbb{Z}\atop a-b=m} [b]_k :B^*_aB_b:
-\rho\sum_{a,b\in\mathbb{Z}\atop a-b=m} [b+1]_k :B^*_aB_b:.  \label{eqn:P-kP}
\end{align}
In the second equality above, we have used identity $[b]_k+ k [b]_{k-1}=[b+1]_k$.

Replacing  $a$ by $a-1$ and $b$ by $b-1$ in the second term on the right hand side
of equation \eqref{eqn:P-kP}, we obtain
\begin{align*}
(1-\rho)P^{(k)}_m-\rho kP^{(k-1)}_m
=&\sum_{a,b\in\mathbb{Z}\atop a-b=m} [b]_k \big(:B^*_aB_b: -\rho:B^*_{a-1}B_{b-1}:\big).
\end{align*}
 Equation \eqref{eqn:(1-rho)O-rho O=...}
then follows from this equation and Lemma \ref{cor::BB:-rho:BB:}. The theorem is thus proved.
$\Box$

For general $\rho$, it is not clear whether the infinite sum on the right hand side of
equation \eqref{eqn:O as BB} could be written as a sum of two infinite summations, one
containing $B_{a}^*B_{b}$ and another one containing $B_{a-1}^*B_{b-1}$,
since each of them may not be well defined.
However, if $\rho=0$ or $\rho = -1$, this will be possible after introducing a new product
\begin{equation}\label{eqn:def star}
B_{a}^* \star B_{b} := B_{a}^*  B_{b} - (1-\rho) \delta_{a,b} \delta_{b \geq 0}.
\end{equation}

If fact, if $\rho=0$, by equation \eqref{eqn:commutation relation for HL:B*B}, we have
\begin{equation} \label{eqn:str0}
 B_{a}^* \star B_{b} = \left\{ \begin{array}{ll}
                                B_{a}^*  B_{b}, & {\rm if} \,\,\,  b<0, \\
                                - B_{b+1} B_{a+1}^*, & {\rm if} \,\,\, b \geq 0.
                                \end{array} \right.
\end{equation}

For $\rho=-1$, the actions of $B_n$ and $B^*_n$ on $\mathbb{C}[\mathbf{\hat{t}}]$ are the same as the actions of
$\hat{B}_n$ and $\hat{B}^*_n$ respectively. Hence by equations \eqref{eqn:BB=-BB+delta} and \eqref{eqn:B*=B rho=-1}, we have
\begin{equation} \label{eqn:str-1*}
 B_{a}^* \star B_{b} = \left\{ \begin{array}{ll}
                                B_{a}^*  B_{b}, & {\rm if} \,\,\,  b<0, \\
                                - B_{b} B_{a}^*, & {\rm if} \,\,\, b \geq 0
                                \end{array} \right.
 \end{equation}
when considered as operators acting on $\mathbb{C}[\mathbf{\hat{t}}]$.

\begin{rem}
	All operators in this paper (including $B_a^*$ and $B_b$) are bosonic operators (i.e., differential operators).
	The notation $``: \,\, :"$ always represents the normal ordering for bosonic operators as explained in equation \eqref{eqn:def::}.
	When $\rho=0$ or $-1$,
	the formula for $B_a^*\star B_b$ in equations \eqref{eqn:str0} and \eqref{eqn:str-1*} are similar to the fermionic normal ordering for charged and neutral fermions (see, for examples, \cite{Alex21,DJM}).
	But in this paper,
	we do not consider $B_a^*$ and $B_b$ as fermionic operators.
	It is convenient to use the $\star$-product notation here to unify these two cases and avoid the confusion with bosonic normal ordering.
	Moreover,
	the $\star$-product defined in equation \eqref{eqn:def star} makes sense for all $\rho\in\mathbb{C}$.
	For general $\rho$,
	we do not know any fermionic interpretation for this notation.
\end{rem}

When $\rho=0$ or $-1$, for any constants $\gamma_{a,b} \in \mathbb{C}$ and $m \in \mathbb{Z}$, the
 infinite summation
  \[ \sum_{a, b \in \mathbb{Z} \atop a-b=m}  \gamma_{a,b} \,\, B_{a}^* \star B_{b} \]
  is well defined  since for any fixed polynomial $f$,
 \[ (B_{a}^* \star B_{b}) \cdot f=0\]
 if $b < - \deg(f)$ or $b \geq 0$ and $a > \deg(f)$.
 Here $f \in \mathbb{C}[\mathbf{t}]$ if $\rho=0$ or
 $f \in \mathbb{C}[\mathbf{\hat{t}}]$ if $\rho=-1$.
 In these cases, equation \eqref{eqn:O as BB} can be further simplified as follows.
\begin{cor} \label{cor:O as BB}
Assume $\rho=0$ or $\rho=-1$.
When considered as operators on $\mathbb{C}[\mathbf{t}]$ if $\rho=0$
or operators on $\mathbb{C}[\mathbf{\hat{t}}]$ if $\rho=-1$, we have
\begin{align}\label{eqn:O=BB}
P^{(k)}_m
= \sum_{a,b\in\mathbb{Z}\atop a-b=m}  g_{k}(b;\rho) \, B_{a}^* \star B_{b} +\delta_{m,0}\frac{\rho^k k!}{(1-\rho)^k},
\end{align}
where
\begin{align} \label{def:g_k}
g_{k}(b;\rho) := - k! \sum_{j=0}^{k-1} \frac{\rho^j }{(1-\rho)^{j+1}}\binom{b}{k-1-j}.
\end{align}
\end{cor}
{\bf Proof}:
 By Theorem \ref{thm:O as BB} and the definition of $B^*_a \star B_b$, we have
\begin{align*}
&P^{(k)}_m - \delta_{m,0}\frac{\rho^k k!}{(1-\rho)^k} \\
=& \sum_{a,b\in\mathbb{Z}\atop a-b=m}  h_{k}(b;\rho) \left\{ B_{a}^* \star B_{b} + (1-\rho) \delta_{a,b} \delta_{b \geq 0}
                - B_{a-1}^* \star B_{b-1} - (1-\rho) \delta_{a,b} \delta_{b \geq 1} \right\} \\
=& \delta_{m,0} (1-\rho) h_k(0; \rho)  + \sum_{a,b\in\mathbb{Z}\atop a-b=m}  h_{k}(b;\rho)  B_{a}^* \star B_{b}
   - \sum_{a,b\in\mathbb{Z}\atop a-b=m}  h_{k}(b;\rho)  B_{a-1}^* \star B_{b-1}.
\end{align*}
Note that $h_k(0; \rho)=0$. Replacing $a$ by $a+1$ and $b$ by $b+1$ for the last term
in the above equation, we obtain
\begin{align*}
P^{(k)}_m - \delta_{m,0}\frac{\rho^k k!}{(1-\rho)^k}
=&  \sum_{a,b\in\mathbb{Z}\atop a-b=m}  \left\{ h_{k}(b;\rho)-h_{k}(b+1;\rho) \right\} B_{a}^* \star B_{b}.
\end{align*}
Using the identity $\binom{b}{k-j}-\binom{b+1}{k-j}=-\binom{b}{k-j-1}$, we have
\[ h_{k}(b;\rho)-h_{k}(b+1;\rho) = - k! \sum_{j=0}^{k-1} \frac{\rho^j }{(1-\rho)^{j+1}}\binom{b}{k-1-j}=g_{k}(b;\rho).\]
We thus obtain the desired formula.
$\Box$

\section{Actions of W-type operators $\widehat{P}^{(k)}_m$ on Q-functions} \label{sec:pf2}

In this section, we prove Theorem \ref{thm:main2} and give some applications of this theorem.
Throughout this section, we set $\rho=-1$ and assume that all differential operators
act on the ring $\mathbb{C}[\mathbf{\hat{t}}]$ where $\mathbf{\hat{t}}=(t_1, t_3, t_5, \cdots)$.
Under these assumptions, actions of operators $B_n$ and $B_n^*$ are the same as
the actions of  $\hat{B}_n$ and $\hat{B}_n^*$. To keep notations simple, we will use
$B_n$ instead of $\hat{B}_n$. In view of equation \eqref{eqn:B*=B rho=-1}, we can
replace $B_n^*$ by $(-1)^n B_{-n}$ in Corollary \ref{cor:O as BB} and obtain the following formula
\begin{align}\label{eqn:O=BBQ}
\widehat{P}^{(k)}_m
= \sum_{a,b\in\mathbb{Z}\atop a+b=-m}  (-1)^{a+1} d_{k}(b) \, B_{a} \star B_{b} +\delta_{m,0}(-2)^{-k} k!,
\end{align}
where $d_k(b)=-g_k(b;-1)$ is given by equation \eqref{eqn:d2} and
\begin{align} \label{eqn:BsBQ}
B_a\star B_b
=B_aB_b-2(-1)^b\delta_{a,-b}\delta_{b\geq0}.
\end{align}

\subsection{Proof of Theorem \ref{thm:main2}}

To prove Theorem \ref{thm:main2}, we need the following two lemmas.
\begin{lem}\label{lem:M on 1}
Assume $\rho=-1$. For any $k\in\mathbb{Z}_{+}$, we have
\begin{align*}
\widehat{P}^{(k)}_m \cdot 1
=\delta_{m<0} \sum_{a=0}^{-m} (-1)^{m-a} d_k(a)Q_{(a,-m-a)}(\mathbf{\hat{t}}).
\end{align*}
\end{lem}
{\bf Proof:}
Equation \eqref{eqn:str-1*} can be rewritten as
\begin{equation} \label{eqn:str-1}
 B_{a} \star B_{b} = \left\{ \begin{array}{ll}
                                B_{a}  B_{b}, & {\rm if} \,\,\,  b<0, \\
                                - B_{b} B_{a}, & {\rm if} \,\,\, b \geq 0
                                \end{array} \right.
 \end{equation}
 for all $a, b \in \mathbb{Z}$. Since $B_n \cdot 1 = \delta_{n, 0}$ if $n \leq 0$, we have
\[\big(B_a\star B_b\big) \cdot 1 =0 \]
if $a<0$ or $b<0$.
By equation \eqref{eqn:O=BBQ}, we have
\begin{align*}
\widehat{P}^{(k)}_m \cdot 1
=&\delta_{m\leq0} \cdot \sum_{a=0}^{-m} (-1)^{a+1} d_k(-m-a) (B_a B_{-m-a}-2\delta_{a,0}\delta_{m,0}) \cdot 1
+\delta_{m,0} (-2)^{-k} k!\\
=&\delta_{m<0} \cdot \sum_{a=0}^{-m} (-1)^{m-a} d_k(a) \cdot Q_{(a,-m-a)}(\mathbf{\hat{t}}).
\end{align*}
In the second equality, we have used the fact that $d_k(0)=\frac{(-1)^{k-1}k!}{2^k}$, $Q_{(0,0)}(\mathbf{\hat{t}})=Q_\emptyset(\mathbf{\hat{t}})=1$,
and equation \eqref{eqn:BB=-BB+delta}.
This lemma is thus proved.
$\Box$

\begin{lem}\label{lem:[BB,B]}
For any $a,b,n\in\mathbb{Z}$, we have
\begin{align*}
[B_a \star B_b ,B_n]
=2(-1)^n(\delta_{n,-b}B_a-\delta_{n,-a}B_b).
\end{align*}
\end{lem}
{\bf Proof:}
 Since $B_a\star B_b - B_a B_b$ is a constant, we have
\begin{align*}
[B_a \star B_b,B_n]
=&[B_a B_b,B_n]
=B_aB_bB_n-B_nB_aB_b\\
=&B_a[B_b,B_n]_+ - [B_a,B_n]_+ B_b\\
=&2(-1)^n(\delta_{n,-b}B_a-\delta_{n,-a}B_b),
\end{align*}
where $[B_a,B_b]_+=B_aB_b+B_bB_a$. In the last equality, we have used equation \eqref{eqn:BB=-BB+delta}.
$\Box$

Now we are ready to prove our first main theorem.

\noindent{\bf Proof of Theorem \ref{thm:main2}:}
By equation \eqref{eqn:O=BBQ} and Lemma \ref{lem:[BB,B]}, we have
\begin{align*}
[\widehat{P}^{(k)}_m , B_n]
= \sum_{a,b\in\mathbb{Z}\atop a+b=-m}  (-1)^{a+1} d_{k}(b)  \cdot 2(-1)^n(\delta_{n,-b}B_a-\delta_{n,-a}B_b)
=c_{k}^{m}(n) B_{n-m},
\end{align*}
where $c_{k}^{m}(n) =2d_k(n-m) -2(-1)^{m}d_k(-n)$.

Since $Q_\la(\mathbf{\hat{t}}) = B_{\la_1}\cdots B_{\la_l}\cdot1$, we can use the above commutation relation repeatedly to obtain
\begin{align*}
\widehat{P}^{(k)}_m \cdot Q_\la(\mathbf{\hat{t}})
=&\widehat{P}^{(k)}_m B_{\la_1}\cdots B_{\la_l}\cdot1\\
=&\sum_{i=1}^{l(\la)} c_k^m(\la_i) B_{\la_1}\cdots B_{\la_i-m} \cdots B_{\la_l}\cdot1
+B_{\la_1}\cdots B_{\la_l} \widehat{P}^{(k)}_m \cdot1.
\end{align*}
The theorem then follows from Lemma \ref{lem:M on 1} and definition of Q-functions. $\Box$

\subsection{Special cases of Theorem \ref{thm:main2}}

As explained at the end of Section \ref{sec:pre},
action of $\widehat{P}^{(k)}_m$ on $\mathbb{C}[\mathbf{\hat{t}}]$ is the
same as the action of $\widehat{M}^{(k)}_m$, whose generating series
$\widehat{M}^{(k)}(z)=\sum_{m \in \mathbb{Z}} \widehat{M}^{(k)}_m z^{-k-m}$ is given by
\[ \widehat{M}^{(k)}(z) = \,\, :(\partial_z +\hat{J}(z))^{k-1} \cdot \hat{J}(z) :,\]
and $\hat{J}(z)$ is defined by equation \eqref{eqn:Jhat}.
Since $Q_\la \in \mathbb{C}[\mathbf{\hat{t}}]$ for all $\la$, we can replace
$\widehat{P}^{(k)}_m$ by $\widehat{M}^{(k)}_m$ in equation \eqref{eqn:main2}.
We observe that $Q_\la$  is an eigenfunction of $\widehat{M}^{(k)}_0$ for all $\la \in \mathbb{Z}^l$ and $k \geq 0$.

\begin{exa}\label{exa:k=1 main}
Consider the $k=1$ case.
Operators in $\widehat{M}^{(1)}(z) = \hat{J}(z)$ generate a Heisenberg algebra.
Equation \eqref{eqn:main2} in this case is equivalent to
\begin{align} \label{eqn:tQ}
mt_m Q_\la
= \sum_{i=1}^l Q_{\la+m\epsilon_i}+ \sum_{a=0}^{m} \frac{(-1)^{m-a}}{4} Q_{(\la,a,m-a)},\ \ \
\end{align}
and
\begin{align} \label{eqn:derQ}
\frac{\partial}{\partial t_m} Q_\la
= 2\sum_{i=1}^l Q_{\la-m\epsilon_i}
\end{align}
for all positive odd integers $m$. These equations are well known. For example,
equation \eqref{eqn:derQ} can be found on page 266 in \cite{Mac}.
Equation \eqref{eqn:tQ} can be found in  \cite{B} and \cite{LY}.

Recall that operators $m t_m$ and $\frac{1}{2} \frac{\partial}{\partial t_m}$ are adjoint to each other with respect to the standard inner product
on $\mathbb{C}[\mathbf{\hat{t}}]$  defined by
\begin{align} \label{eqn:inner}
\langle Q_\lambda,Q_\mu\rangle=2^{l(\lambda)}\delta_{\lambda,\mu}
\end{align}
 when $\lambda, \mu$ are strict partitions
(see Section III.8 in \cite{Mac}). More precisely, let $f^{\perp}$ be the adjoint of an operator $f$ on $\mathbb{C}[\mathbf{\hat{t}}]$
with respect to the above inner product. Then
\begin{align} \label{eqn:adjoint}
(t_m)^{\perp} = \frac{1}{2m}\frac{\partial}{\partial t_m}
\end{align}
for all odd positive integers $m$. This equation will be used in Sections \ref{sec:pf2app} and \ref{sec:pf KW}.

\end{exa}

\begin{exa} \label{exm:VirQ}
Consider the $k=2$ case.
Note that  $\widehat{M}^{(2)}(z)=\,\,:\hat{J}(z)^2+\partial_z \hat{J}(z):$.
Let $\hat{L}_m$ be the coefficient of $z^{-2m-2}$ in
\[   \frac{1}{2} :\hat{J}(z)^2: \,\,=\,\, \frac{1}{2} \left( \widehat{M}^{(2)}(z)-  \partial_z \widehat{M}^{(1)}(z) \right). \]
Since $\partial_z \widehat{M}^{(1)}(z)$ only contains $z^n$ with $n$ odd, it does not contribute to $\hat{L}_m$.
In fact
$$\hat{L}_m = \frac{1}{2} \widehat{M}^{(2)}_{2m}$$
 for all integers $m$. They can be written explicitly as
\begin{align*}
\hat{L}_m
= & \sum_{a>\max\{0,-2m\}\atop a\in\mathbb{Z}_{odd}} 2at_a\frac{\partial}{\partial t_{a+2m}}
+\frac{\delta_{m>0}}{2}\sum_{k=1\atop k\in\mathbb{Z}_{odd}}^{2m-1} \frac{\partial^2}{\partial t_k \partial t_{2m-k}}
+ 2\delta_{m<0}\sum_{k=1\atop k\in\mathbb{Z}_{odd}}^{-2m-1} k(-2m-k)t_k t_{-2m-k}.
\end{align*}
These operators generate a Virasoro algebra.
As a consequence of equation \eqref{eqn:main2}, we obtain
\begin{align} \label{eqn:VirQ}
\hat{L}_m \cdot Q_\la
=\sum_{i=1}^l 2(\la_i - m) Q_{\la-2m\epsilon_i}
+ \delta_{m<0}\sum_{a=0}^{-2m} \frac{(-1)^{a}(2a-1)}{4} Q_{(\la,a,-2m-a)}
\end{align}
for all integers $m$. If $m>0$, this formula was proved in \cite{ASY}, where the notations are slightly different from what we are using here.
The formulas for $\hat{L}_{-1} \cdot Q_\la$ and $\hat{L}_{-2} \cdot Q_\la$  were established in \cite{LY} and \cite{LY2}.
Note that by the Virasoro relation
\[[\hat{L}_{n}, \hat{L}_{m}]=4(n-m) \hat{L}_{m+n} \]
 for $n, m < 0$,
$\hat{L}_{-1}$ and $\hat{L}_{-2}$ generate $\hat{L}_{m}$ for all $m<0$.
Another proof for $\hat{L}_{m} Q_\la$ for all $m<0$ was given in \cite{ASY2}. A more general formula for actions of Virasoro operators
 on Hall-Littlewood functions with $\rho^n=1$ for any $n \geq 2$ was given in \cite{LY3}.

Equation \eqref{eqn:VirQ} for $-2 \leq m \leq 2$ was used in \cite{LY} and \cite{LY2} to prove the Q-function
expansion formulas for Kontsevich-Witten and BGW tau-functions.
\end{exa}

\begin{exa}\label{exa:main k=3}
Consider the $k=3$ case. The coefficients in equation \eqref{eqn:main2} are given by
\begin{align*}
d_3(a)=&\frac{3a^2}{2}-3a+\frac{3}{4},\\
c_3^m(n)=& \left\{
\begin{array}{ll}
3(m^2-2mn+2m+2n^2+1),\,\,\, & {\rm if} \,\,\,  m \,\,\, {\rm is \,\,\,  odd},\\
3(m+2)(m-2n), \,\,\, & {\rm if} \,\,\,  m \,\,\, {\rm is \,\,\,  even}.
\end{array} \right.
\end{align*}
Let $\widehat{W}_m$ be the coefficient of $ z^{-m-3}$ in
\[ \frac{1}{3}:\hat{J}(z)^3: \,\,=\,\, \frac{1}{3} \widehat{M}^{(3)}(z) - \frac{1}{2} \partial_z \widehat{M}^{(2)}(z)
            + \frac{1}{6} \partial_z^2 \widehat{M}^{(1)}(z). \]
Then
\begin{align}\label{eqn:W^B_m}
\widehat{W}_m\,:=& \, \frac{1}{3} \widehat{M}^{(3)}_m + \frac{m+2}{2} \widehat{M}^{(2)}_m+\frac{(m+1)(m+2)}{6} \widehat{M}^{(1)}_m
\end{align}
for odd integers $m$.
By equation \eqref{eqn:main2}, we have
\begin{align}
\widehat{W}_m \cdot Q_\la
=&\sum_{i=1}^l \left( 2\la_i^2-2m\la_i+\frac{(m+1)(m-1)}{3} \right) Q_{\la-m\epsilon_i}  \nonumber \\
&+ \delta_{m<0}\sum_{a=0}^{-m} (-1)^{a+1} \left(\frac{a^2}{2}+\frac{ma}{2}+\frac{m^2}{12}-\frac{1}{12}\right)Q_{(\la,a,-m-a)}
                            \label{eqn:WQ}
\end{align}
for all odd integers $m$.

Operators $\widehat{W}_{-1}$ and $\widehat{W}_{\pm 3}$ are particularly interesting since they are related to the cut-and-join operators
for Br\'{e}zin-Gross-Witten and Kontsevich-Witten  tau-functions respectively. As applications, we will use
formula \eqref{eqn:WQ} with $m=-1$ and $m=\pm 3$ to give new proofs for Alexandrov's conjecture and Mironov-Morozov formula in Section \ref{sec:pf2app} and Section \ref{sec:pf KW} respectively.

\end{exa}

\begin{exa}\label{exa:main k=4}
Consider the $k=4$ case.
Let $\widehat{N}_m$ be the coefficient of $z^{-m-4}$ in
\[ \frac{1}{4} \left( :\hat{J}(z)^4-(\partial_z \hat{J}(z))^2: \right)
\,\,\, = \,\,\, \frac{1}{4} \left( \widehat{M}^{(4)}(z) - 2 \partial_z \widehat{M}^{(3)}(z) + \partial_z^2 \widehat{M}^{(2)}(z) \right)\]
for even integers $m$.
Note that $:\hat{J}(z)^4:$ is not a linear combination of $\partial_z^n \widehat{M}^{(4-n)}(z)$ with $0 \leq n \leq 4$.

By equation \eqref{eqn:main2}, we have
\begin{align*}
\widehat{N}_m \cdot Q_\la
=&\sum_{i=1}^l (2\la_i^3-3m\la_i^2+m^2\la_i-m\la_i-2\la_i+m^2/2+m) Q_{\la-m\epsilon_i} \\
&+ \delta_{m<0}\sum_{a=0}^{-m} \frac{(-1)^m}{8} \big(m^2(2a-1)+2m(3a^2-a-1)+4a^3-4a\big) Q_{(\la,a,-m-a)},
\end{align*}
for all even integers $m$.
\end{exa}

\subsection{Application to BGW model} \label{sec:pf2app}

The BGW tau-function, denoted by $\tau_{BGW}$, was originally introduced by Br\'{e}zin, Gross, and Witten for studying lattice gauge theory (see \cite{BG} and \cite{GW}). It also has enumerative geometric interpretations as the generating function
 of certain intersection numbers on moduli spaces of stable curves (see \cite{N17} \cite{KN} \cite{YZ}).
 The cut-and-join description for $\tau_{BGW}$ was given in \cite{Alex18}. The following formulation
  of the cut-and-join description was given by equation (2.27) in \cite{Alex20}:
\begin{align}\label{eqn:caj expression of BGW}
\tau_{BGW}(\mathbf{\hat{t}}) =e^{\hbar W_{BGW}}\cdot1,
\end{align}
where $\hbar$ is a formal parameter, and
\begin{align*}
W_{BGW}=\frac{1}{4} \widehat{W}_{-1}+\frac{1}{8}t_1
\end{align*}
is called the cut-and-join operator, $\widehat{W}_{-1}$ is defined by equation \eqref{eqn:W^B_m} with $m=-1$.
A more explicit description for $\widehat{W}_{-1}$ is the following
\begin{align*}
\widehat{W}_{-1}=&\sum_{a,b\in\mathbb{Z}_{+,odd}} \Big(4abt_at_b\frac{\partial}{\partial t_{a+b-1}}
+2(a+b+1)t_{a+b+1}\frac{\partial^2}{\partial t_a \partial t_b}\Big).
\end{align*}

In this subsection, we use equation \eqref{eqn:WQ} to give a new simple combinatorial proof of the following theorem which
was conjectured by Alexandrov in \cite{Alex20} and has been proved independently in
\cite{Alex21} and \cite{LY2} using different methods.

\begin{thm}\label{thm:BGW schur Q} (Alexandrov, Liu-Yang)
\begin{align*}
	\tau_{BGW}(\mathbf{\hat{t}})
	=&\sum_{\la\in DP} 2^{-l(\la)} \left( \frac{\hbar}{16} \right)^{|\la|}\frac{E_\la^3}{E_{2\la}^2} \cdot
        Q_\la(\mathbf{\hat{t}}),
\end{align*}
	where DP denotes the set of all strict partitions (i.e. partitions with positive distinct parts),
    $l(\la)$ is the number of parts in $\la$, $|\la|$ is the sum of all parts of $\la$,
    $2(\lambda_1,...,\lambda_{l(\lambda)})
    :=(2\lambda_1,...,2\lambda_{l(\lambda)})$,
    and
	$E_\la:=Q_\la(1, 0, 0, \cdots)$ is a constant depending on $\la$.
\end{thm}

\begin{rem}
It is well known that
\begin{align} \label{eqn:A/A}
E_\la = \frac{2^{|\la|}}{\la_1 ! \cdots \la_l !} \prod_{i<j} \frac{\la_i-\la_j}{\la_i+\la_j},
  \hspace{10pt} {\rm and} \hspace{10pt}
\frac{E_\la}{E_{2\la}}
=\prod_{i=1}^{l}(2\la_i-1)!!
\end{align}
for any strict partition $\la=(\la_1, \cdots, \la_l)$ (see, for example, \cite{Alex20}).
\end{rem}

\vspace{6pt}
\noindent{\bf Proof of Theorem \ref{thm:BGW schur Q}:}
By equation \eqref{eqn:WQ} with $m=-1$ and equation \eqref{eqn:tQ} with $m=1$, we have
\begin{align*}
   W_{BGW} \cdot Q_\la
     =& \frac{1}{8} \sum_{i=1}^{l} (2 \la_i+1)^2 Q_{\la+\epsilon_i} + \frac{1}{16} Q_{(\la, 1)}
\end{align*}
for all $\la \in \mathbb{Z}^l$. When $\la$ is a strict partition, this formula can be written as
\begin{align}\label{eqn:W^BGW action}
	W_{BGW} \cdot Q_\la
	=& \frac{1}{8} \sum_{\mu=\la+\square \in DP} \frac{(2i_\mu(\square)-1)^2}{2^{l(\mu)-l(\la)} } \, Q_\mu.
\end{align}
In this formula, we have identified a partition with its Young diagram (also called Ferrers diagram, see \cite{S}). The notation
$\mu=\la+\square$ means that the Young diagram of $\mu$ is obtained from the Young diagram of $\la$ by adding one box from the right such that the resulting diagram still corresponds to a strict partition.
In case $\mu$ is no longer strict, $Q_\mu = 0$ and the corresponding term does not appear in this formula.
$i_\mu(\square)$ is the horizontal coordinate of this newly added box $\square$, i.e. $i_\mu(\square)=\mu_i=\la_i+1$ if
$\square$ is added to the $i$-th row of the Young diagram of $\la$ for $1\leq i\leq l(\la)$,
and $i_{\mu}(\square)=1$ if $\square$ is added to the bottom of the Young diagram of $\la$ (in this case $\mu=(\la,1)$ ).

Repeatedly applying equation \eqref{eqn:W^BGW action} to $Q_\emptyset(\mathbf{\hat{t}})=1$, we obtain the following
formula
\begin{align}\label{eqn:W^n-action}
(W_{BGW})^n \cdot 1
=\sum_{\mu\in DP\atop |\mu|=n}
\frac{ c(\mu)\Big(\prod_{i=1}^{l(\mu)}(2\mu_i-1)!!\Big)^2} {2^{l(\mu)}\cdot8^{n}} \cdot Q_\mu(\hat{\mathbf{t}})
\end{align}
for all $n \geq 0$, where $c(\mu)$ is the number of sequences of strict partitions $(\mu^1, \cdots, \mu^n)$
such that $ |\mu^i|=i$ for each $i$, and
\[\emptyset \subset \mu^1 \subset \dots \subset \mu^n=\mu. \]
In other words, $c(\mu)$ is the number of ways to obtain $\mu$ from empty partition by adding $n=|\mu|$ boxes one by one while keeping
each partition in this process to be strict.  It is well known that this number is
related to the value of Green's polynomial associated to partitions $\mu$ and $1^n$. Let $n=|\mu|$. By
 part (c) of example 11 in Section III.8  in \cite{Mac},
\[c(\mu) = \left\langle Q_\mu, t_1^n \right\rangle
=\frac{1}{2^n} \left\langle \frac{\partial^n}{\partial t_1^n} Q_\mu, \, 1 \right\rangle,\]
where the last equality follows from equation \eqref{eqn:adjoint}.
Since $Q_\mu$ is a homogeneous polynomial of degree $n$, we can write
\[ Q_{\mu}(\mathbf{\hat{t}})= a t_1^n + \sum_{k=0}^{n-1} t_1^k f_k(t_3, t_5, \cdots ) \]
where $a$ is a constant and $f_k(t_3, t_5, \cdots )$ is a homogeneous polynomial of degree $n-k$.
In particular $E_\mu=Q_\mu(1, 0, 0, \cdots)=a$ and  $\frac{\partial^n}{\partial t_1^n} Q_\mu= n! a = n! E_\mu$.
Hence we have
\[ c(\mu)=\frac{n! }{2^{n}} E_\mu. \]
Therefore, by equations \eqref{eqn:caj expression of BGW} and \eqref{eqn:W^n-action}, we have
\begin{align*}
\tau_{BGW}(\mathbf{\hat{t}}) = \sum_{n=0}^\infty \frac{\hbar^n}{n!} (W_{BGW})^n \cdot 1
=&\sum_{\mu\in DP}
\frac{ \hbar^{|\mu|} E_\mu \Big(\prod_{i=1}^{l(\mu)}(2\mu_i-1)!!\Big)^2} {2^{l(\mu)}\cdot 16^{|\mu|}} \cdot Q_\mu(\mathbf{\hat{t}}).
\end{align*}
Together with equation \eqref{eqn:A/A}, this proves the theorem.
$\Box$

\subsection{Application to Kontsevich-Witten model} \label{sec:pf KW}
Kontsevich-Witten tau-function $\tau_{KW}(\mathbf{\hat{t}})$ is the generating function for intersection numbers of $\psi$-classes on moduli
spaces of stable curves. It is also the partition function for $2d$ topological gravity (see \cite{W90}, \cite{Ko}).
The cut-and-join operator for Kontsevich-Witten tau-function was introduced in \cite{Alex11}. Following the formulation
given in equation (2.27) in \cite{Alex20}, this operator can be written as
\begin{align*}
	W_{KW}=\frac{1}{12} \widehat{W}_{-3} - \frac{t_1^3}{18} +\frac{t_3}{8},
\end{align*}
where $\widehat{W}_{-3}$ is defined by equation \eqref{eqn:W^B_m} with $m=-3$. More explicitly,
\begin{align*}
	\widehat{W}_{-3}=&\sum_{a,b\in\mathbb{Z}_{>0,odd}} \Big(4abt_at_b\frac{\partial}{\partial t_{a+b-3}}
	+2(a+b+3)t_{a+b+3}\frac{\partial^2}{\partial t_a \partial t_b}\Big)
	+\frac{8t_1^3}{3}.
\end{align*}

\noindent The cut-and-join equation for Kontsevich-Witten tau-function is
\begin{align}\label{eqn:caj eq for KW}
\frac{\partial}{\partial \hbar} \tau_{KW}
- W_{KW} \cdot \tau_{KW} = 0,
\end{align}
which uniquely specifies $\tau_{KW}$ as a formal power series with initial values $\tau_{KW}|_{\hbar=0}=1$
(see equation (2.25) in \cite{Alex20}).

In this subsection, we will use equation \eqref{eqn:WQ} to give a new proof to the following theorem:
\begin{thm}\label{thm:KW Schur Q} (Mironov-Morozov \cite{MM20}, Liu-Yang \cite{LY})
\begin{align}\label{eqn:KW Schur Q}
\tau_{KW}(\mathbf{\hat{t}})
=&\sum_{\la\in DP} 2^{-l(\la)} \left( \frac{\hbar}{16} \right)^{|\la|/3} \frac{E_\la}{E_{2\la}} \cdot A_{2\la} \cdot
Q_\la(\mathbf{\hat{t}}),
\end{align}
where $A_\la=Q_\la(0, 1/3, 0, \cdots)$
and we also follow the notations in Theorem \ref{thm:BGW schur Q}.
\end{thm}
Formula \eqref{eqn:KW Schur Q} was proposed by Mironov and Morozov in \cite{MM20} based on a series of results related to matrix models and Q-functions (see \cite{MM20}, \cite{DIZ}, \cite{J}, \cite{L}, as well as \cite{Alex20}). A proof of this formula using Virasoro constraints was given in \cite{LY}, which does not rely on matrix models. The proof in this subsection also does not need matrix models
and is much simpler than that in \cite{LY}.

To prove Theorem \ref{thm:KW Schur Q}, we first reduce it to a combinatorial identity.
\begin{pro} \label{pro:Phi}
For any $\mu=(\mu_1, \cdots, \mu_l) \in \mathbb{Z}^l$, let
\begin{align} \label{eqn:Phi}
\Phi(\mu):=&
3|\mu| A_{2\mu}
- 6\sum_{i=1}^{l} (2\mu_i-1)(2\mu_i-5) D_i^3 A_{2\mu} +24\sum_{i,j=1\atop i\neq j}^{l(\mu)} D_i^2 D_{j} A_{2\mu}
+8\sum_{i,j,k=1\atop i\neq j\neq k}^{l} D_i D_j D_k A_{2\mu},
\end{align}
where $D_i$ should be viewed as an operator acting on $A_{\mu}$ in the following way
\begin{align} \label{eqn:Di}
D_i A_{\mu} := \frac{A_{\mu-2\epsilon_i}}{\mu_i-1}.
\end{align}
Theorem \ref{thm:KW Schur Q} holds if $\Phi(\mu)=0$ for all strict
 partitions $\mu$.
\end{pro}
{\bf Proof}:
Let $\tau_{MM}(\mathbf{\hat{t}})$ be the function on the right hand side of equation \eqref{eqn:KW Schur Q}.
To prove Theorem~\ref{thm:KW Schur Q}, it suffices to show that $\tau_{MM}$ also satisfies the cut-and-join equation \eqref{eqn:caj eq for KW}. For any $\mu \in \mathbb{Z}^l$, let
\begin{align} \label{eqn:Phitilde}
\widetilde{\Phi}(\mu) :=&
\left\langle Q_\mu, \,\, W_{KW} \cdot \tau_{MM}  - \frac{\partial}{\partial \hbar} \tau_{MM} \right\rangle  \nonumber \\
=& \left\langle (W_{KW})^\perp \cdot Q_\mu, \,\, \tau_{MM} \right\rangle - \left\langle Q_\mu, \,\, \frac{\partial}{\partial \hbar} \tau_{MM} \right\rangle,
\end{align}
where $(\cdot)^\perp$ means taking adjoint with respect to the standard inner product given by
equation \eqref{eqn:inner}.
Since $\{Q_\mu \mid \mu \in DP \}$ is a basis of $\mathbb{C}[\mathbf{\hat{t}}]$, Theorem~\ref{thm:KW Schur Q} holds if
$ \widetilde{\Phi}(\mu) = 0$
for all strict partitions $\mu$.

To compute $\widetilde{\Phi}(\mu)$, we first compute action of $(W_{KW})^\perp$ on Q-functions.
By equation~\eqref{eqn:adjoint},
\begin{align*}
(\widehat{W}_{-3})^\perp=\widehat{W}_3
=\sum_{a,b\in\mathbb{Z}_{>0,odd}} \Big(4abt_at_b\frac{\partial}{\partial t_{a+b+3}}
+2(a+b-3)t_{a+b-3}\frac{\partial^2}{\partial t_a \partial t_b}\Big)
+\frac{1}{3}\frac{\partial^3}{\partial t_1^3}.
\end{align*}
Hence
\begin{align*}
(W_{KW})^\perp
=\frac{1}{12}\widehat{W}_{3}
- \frac{1}{144}\frac{\partial^3}{\partial t_1^3}
+ \frac{1}{48}\frac{\partial}{\partial t_3}.
\end{align*}
By equations \eqref{eqn:derQ} and \eqref{eqn:WQ},  we have
\begin{align}\label{eqn:W_KW^perp-action}
(W_{KW})^\perp &\cdot Q_\mu
=\sum_{i=1}^{l(\mu)} \frac{(2\mu_i-1)(2\mu_i-5)}{24} Q_{\mu-3\epsilon_i}
-\frac{1}{6}\sum_{i,j=1\atop i\neq j}^{l(\mu)} Q_{\mu-2\epsilon_i-\epsilon_j}
-\frac{1}{18}\sum_{i,j,k=1\atop i\neq j\neq k}^{l(\mu)} Q_{\mu-\epsilon_i-\epsilon_j-\epsilon_k}.
\end{align}
Note that this formula is much simpler than the formula for $W_{KW} \cdot Q_\mu$. Also note that even if
$\mu$ is a strict partition, the right hand side of equation \eqref{eqn:W_KW^perp-action} contains
$Q_\nu$ where $\nu$ may not be strict partitions.

For any strict partition $\la=(\la_1, \cdots, \la_l)$, define
\begin{equation} \label{eqn:a-nu}
a_\la := A_{2 \la} \left(  \frac{\hbar}{16} \right)^{|\la|/3} \prod_{i=1}^l (2 \la_i -1)!! .
\end{equation}
Then by equation \eqref{eqn:A/A}, we have
\[ \tau_{MM}(\mathbf{\hat{t}}) = \sum_{\la  \in DP} a_\la \, 2^{-l(\la)} Q_{\la} (\mathbf{\hat{t}}). \]
The definition of $a_\la$ in equation \eqref{eqn:a-nu} makes sense for all $\la \in \mathbb{Z}^l$ if we define
$(-1)!! := 1$ and
\[
  (-2k-1)!! \, := \, \frac{(-1)^{k}}{(2k-1)!!}
\]
  for all  positive integers $k$. With this definition of double factorial, we have
\begin{equation} \label{eqn:dfac}
 (2m + 1)!! = (2m+1) \cdot (2m-1)!!
\end{equation}
for all integers $m$.

By Lemma 4.1 in \cite{LY}, we have
\begin{align}\label{eqn:covariation}
\langle Q_{\nu} , \, \tau_{MM} \rangle= a_\nu,
\hspace{10pt} {\rm \,\,\, and \,\,\,}
\left\langle Q_{\nu} , \, \frac{\partial}{\partial \hbar} \tau_{MM} \right\rangle
= \frac{|\nu|}{3 \hbar} a_\nu
\end{align}
for all $\nu \in \mathbb{Z}^l$ with non-negative components.
Using equation \eqref{eqn:BB=-BB+delta} and the fact $B_{n} \cdot 1 = \delta_{n,0}$ if $n \leq 0$, it is straightforward to
check that equation \eqref{eqn:covariation} also holds for all $\nu \in \mathbb{Z}^l$.

Combining equations \eqref{eqn:Phitilde}, \eqref{eqn:W_KW^perp-action}, and \eqref{eqn:covariation}, we have
\begin{align*}
\widetilde{\Phi}(\mu)
=&-\frac{|\mu|}{3\hbar} a_{\mu} + \sum_{i=1}^{l(\mu)} \frac{4\mu_i^2-12\mu_i+5}{24} a_{\mu-3\epsilon_i}
  - \frac{1}{6}\sum_{i,j=1\atop i\neq j}^{l(\mu)} a_{\mu-2\epsilon_i-\epsilon_j}
  - \frac{1}{18}\sum_{i,j,k=1\atop i\neq j\neq k}^{l(\mu)} a_{\mu-\epsilon_i-\epsilon_j-\epsilon_k}.
\end{align*}
By equations \eqref{eqn:a-nu} and \eqref{eqn:dfac},
\begin{align*}
\widetilde{\Phi}(\mu) =& - \frac{1}{144} \left(  \frac{\hbar}{16} \right)^{|\mu|/3-1} \, \prod_{i=1}^{l(\mu)} (2 \mu_i -1)!! \cdot \Phi(\mu)
\end{align*}
where $\Phi(\mu)$ is defined by equation \eqref{eqn:Phi}.
Hence $\widetilde{\Phi}(\mu)=0$ if and only if $\Phi(\mu)=0$.
The proposition is thus proved.
$\Box$

Let $\widetilde{DP}$ be the set of
all $\mu=(\mu_1, \cdots, \mu_l) \in \mathbb{Z}^l$ with $l$ even and
$\mu_1 > \mu_2 > \cdots > \mu_l \geq 0$.
If $\mu \in DP$ has odd length, then $\tilde{\mu}=(\mu, 0) \in \widetilde{DP}$
and
 $\Phi(\mu)=\Phi(\tilde{\mu})$.
Hence by Proposition \ref{pro:Phi}, to prove Theorem \ref{thm:KW Schur Q}, we only need to show
$\Phi(\mu) = 0$ for all $\mu \in \widetilde{DP}$.
In the rest part of this subsection, we will prove this
 by induction on $l(\mu)$.
This induction process depends on a recursion formula for $A_\la$.

We call a vector $\la = (\la_1, \cdots \la_l) \in \mathbb{Z}^l$ {\it weakly positive} if all $\la_i$ are non-negative and at most
one of them is $0$. If $\la$ is weakly positive of even length, $Q_\la$ can also be  defined by the Pfaffian of a skew
symmetric matrix (see Section III.8 in \cite{Mac}). It follows from properties of Pfaffian that
$A_\la$ satisfies the following recursion relation: For any fixed $1\leq n \leq l$,
\begin{equation} \label{eqn:A rec}
A_{\lambda} = (-1)^{n}\sum_{m=1\atop m\neq n}^{l} (-1)^{m+\delta_{m<n}} A_{(\lambda_m, \lambda_n)} A_{\lambda^{\{n,m\}}},
\end{equation}
where $\delta_{m<n}:=1$ if $m<n$ and $\delta_{m<n}:=0$ if $m \geq n$, and
\begin{equation*}
\la^{\{i_1, \cdots, i_n\}}:=(\la_1, \cdots, \widehat{\la}_{i_1}, \cdots, \widehat{\la}_{i_n}, \cdots, \la_l)
\end{equation*}
is obtained from $\la$ by deleting components $\la_{i_1}, ..., \la_{i_n}$
(see, for example, equation (2.4) in \cite{O} and equation (53) in \cite{LY}).
Equation \eqref{eqn:A rec} is called the {\it expansion of $A_\la$ with respect to the $n$-th component of $\la$}.

{\bf Remark}: 
For $\mu \in \widetilde{DP}$, the right hand side of equation \eqref{eqn:Phi} may contain
$A_\la$ with $\la$ not weakly positive. In fact, all such $\la=(\la_1, \cdots, \la_l)$ have the property
that $\la_i$ are even integers  and
$\la_{i} \geq 2(\mu_l+l-i)-6$ for $1 \leq i \leq l$.
Using Lemma 5.1 in \cite{LY} and equation \eqref{eqn:BB=-BB+delta}, it is straightforward to check that
equation \eqref{eqn:A rec} with $n=l$ also holds for such $\la$.

To prove $\Phi(\mu)=0$ for $\mu \in \widetilde{DP}$, we also need the following properties of operators $D_i$ defined by
equation \eqref{eqn:Di}. For any $\la \in \mathbb{Z}^l$, define
\begin{align} \label{eqn:DA}
D A_{\la} := \sum_{i=1}^{l} D_i A_{\la}.
\end{align}
The following lemma shows that operator $D$ is well behaved with respect to the recursion relation \eqref{eqn:A rec}.
\begin{lem}\label{lem:i neq l-1}
For any $\nu=(\nu_1, \cdots, \nu_l) \in \widetilde{DP}$ with $l \geq 4$, we have 
\begin{align} \label{eqn:DerD}
D A_{2\nu}
=& \sum_{j=1}^{l-1} (-1)^{j+1} \big(A_{(2\nu_{j},2\nu_{l})} \cdot D A_{2\nu^{\{j,l\}}}
+DA_{(2\nu_j,2\nu_{l})} \cdot A_{2\nu^{\{j,l\}}}\big),
\end{align}
and
\begin{align}
D^2 A_{2\nu} \label{eqn:DerDD}
=& \sum_{j=1}^{l-1} (-1)^{j+1} \big( D^2 A_{(2\nu_{j},2\nu_{l})} \cdot A_{2\nu^{\{j,l\}}}
+2D A_{(2\nu_{j},2\nu_{l})} \cdot D A_{2\nu^{\{j,l\}}}
+A_{(2\nu_{j},2\nu_{l})} \cdot D^2 A_{2\nu^{\{j,l\}}}\big).
\end{align}
If in addition $\nu_l \geq 1$, we have
\begin{align} \label{eqn:DerD/2}
D A_{2\nu}
=&\frac{1}{2} \sum_{i, j=1 \atop j\neq i}^{l} (-1)^{i+j+\delta_{i<j}} D A_{(2\nu_i,2\nu_j)} \cdot A_{2\nu^{\{i,j\}}}.
\end{align}
If in addition $\nu_l \geq 2$, we have
\begin{align} \label{eqn:DerDD/2}
D^2 A_{2\nu}
=&\frac{1}{2}\sum_{i,j=1 \atop i \neq j}^{l} (-1)^{i+j+\delta_{i<j}} 
  \big(D^2 A_{(2\nu_i,2\nu_j)} \cdot A_{2\nu^{\{i,j\}}}
       + D A_{(2\nu_i,2\nu_j)} \cdot D A_{2\nu^{\{i,j\}}} \big).
\end{align}
\end{lem}
{\bf Proof:}
By definition,
\begin{equation*} %\label{eqn:D+DD}
D A_{2\nu} = \sum_{a=1}^l D_a A_{2\nu}, \hspace{20pt}  
D^2 A_{2\nu} = \sum_{a, b=1}^l D_a D_b A_{2\nu}.
\end{equation*}
Equations \eqref{eqn:DerD} and \eqref{eqn:DerDD} are obtained by expanding $D_a A_{2\nu}$ and $D_a D_b A_{2\nu}$
in above formulas using equation \eqref{eqn:A rec} with $n=l$ (note that $l$ is even by assumption). Although  $2\nu-2 \epsilon_a$ and $2\nu - 2 \epsilon_a - 2\epsilon_b$
may not be weakly positive, such expansions with respect to the last components are still valid as explained in the remark
after equation \eqref{eqn:A rec}.

If $\nu_l \geq 1$ (i.e. $\nu$ is strict), then $2\nu-2 \epsilon_a$ is weakly positive for all $1 \leq a \leq l$.
So we can always expand each $D_a A_{2\nu}$ in $D A_{2\nu}$ using equation \eqref{eqn:A rec} with $n=a$. 
Equation \eqref{eqn:DerD/2} then follows from the fact 
\[ D_1 A_{(2\nu_a, 2 \nu_m)} = - D_2 A_{(2\nu_m, 2 \nu_a)} \]
for all $m \neq a$, which in turn follows from skew symmetry of $A_{(\la_1, \la_2)}$ with respect to $\la_1$ and $\la_2$
when $(\la_1, \la_2)$ is weakly positive.

Similarly, if $\nu_l \geq 2$, then $2\nu - 2 \epsilon_a - 2 \epsilon_b$ is weakly positive for all $1 \leq a, b \leq l$.
So we can expand each $D_a D_b A_{2\nu}$ in $D^2 A_{2\nu}$ using equation~\eqref{eqn:A rec} with $n=a$ and obtain
equation~\eqref{eqn:DerDD/2} using symmetry.
The lemma is thus proved.
$\Box$

\begin{lem} \label{lem:Gamma}
For any $\nu=(\nu_1, \cdots, \nu_l) \in \mathbb{Z}^l$ with even length $l \geq 4$, define
\begin{align} \label{eqn:Gamma}
\Gamma(\nu)
:=\sum_{i,j=1 \atop i\neq j}^{l} (-1)^{i+j+\delta_{j>i}}D A_{(2\nu_i,2\nu_j)} \cdot DA_{2\nu^{\{i,j\}}} .
\end{align}
Then 
$ \Gamma(\nu)=0 $
for all $\nu \in \widetilde{DP}$ with $l(\nu) \geq 4$.
\end{lem}
{\bf Proof}:
We prove this lemma by induction on $l=l(\nu)$. 

Recall $A_{\mu}=0$ if $|\mu|/3$ is not an integer (see equation (32) in \cite{LY}) and
\begin{eqnarray}
A_{(3k_1, 3k_2)} &=& \left( \frac{2}{3} \right)^{k_1+k_2} \frac{1}{k_1 ! k_2 !} \, \, \frac{k_1-k_2}{k_1+k_2}, \label{eqn:Ap2}\\
A_{(3m+1, 3n+2)} &=&  \left( \frac{2}{3} \right)^{m+n+1} \frac{2}{m ! n !(m+n+1)}. \label{eqn:Al<3}
\end{eqnarray}
for $k_1, k_2,m,n \geq 0$ and $(k_1, k_2) \neq (0, 0)$ (see equations (30) and (31) in \cite{LY}).
Using these formulas, it is straightforward to show
\begin{equation} \label{eqn:Dmod2}
D A_{2(3m+2, 3n+2)} = 0
\end{equation}
for all $m, n \geq 0$, and
$\Gamma(\nu) =0$ if $l=4$ and $\nu_i \geq 0$ for all $i$.

Assume $l \geq 6$ and $\nu \in \widetilde{DP}$. 
We will expand $D A_{2\nu^{\{i,j\}}}$ on the right hand side of equation \eqref{eqn:Gamma} according to the following rules:
We apply equation \eqref{eqn:DerD} to $D A_{2\nu^{\{i,j\}}}$ if both $i, j < l$ and apply
 equation \eqref{eqn:DerD/2} to $D A_{2\nu^{\{i,j\}}}$ if $i$ or $j$ is equal to $l$.
 We then obtain the following equation
\begin{align*}
\Gamma(\nu)
=\sum_{k=1}^{l(\nu)-1} (-1)^{k+1} A_{(2\nu_k,2\nu_l)} \cdot \Gamma(\nu^{\{k,l\}})
+\sum_{i,j,k=1 \atop i<j<k}^{l(\nu)-1}
(-1)^{i+j+k}
\Gamma((\nu_i,\nu_j,\nu_k,\nu_l)) \cdot A_{2\nu^{\{i,j,k,l\}}}&.
\end{align*}
Hence $\Gamma(\nu)=0$ by induction.
The lemma is thus proved.
$\Box$

\begin{lem} \label{lem:Psi}
For any $\nu=(\nu_1, \cdots, \nu_l) \in \mathbb{Z}^l$ with even length $l \geq 4$, define
\begin{align}\label{eqn:def Psi}
\Psi(\nu) :=\sum_{i=1}^{l-1}(-1)^{i+1}
\Big( DA_{(2\nu_i,2\nu_l)} \cdot D^2 A_{2\nu^{\{i,l\}}}
+D^2 A_{(2\nu_i,2\nu_l)} \cdot D A_{2\nu^{\{i,l\}}} \Big).
\end{align}
Then $\Psi(\nu) =0$ for all $\nu \in \widetilde{DP}$ with $l(\nu) \geq 4$.
\end{lem}
{\bf Proof}:
We prove this lemma by induction on $l=l(\nu)$. If $l=4$, it is straightforward to check $\Psi(\nu) =0$
using equations
\eqref{eqn:Ap2}, \eqref{eqn:Al<3}, and \eqref{eqn:Dmod2}.

Assume $l \geq 6$. We will expand the right hand side of equation \eqref{eqn:def Psi} according to the following rules:
For $i < l-1$,  we apply equation \eqref{eqn:DerD} to $D A_{2\nu^{\{i,l\}}}$ and apply
 equation~\eqref{eqn:DerDD} to $D^2 A_{2\nu^{\{i,l\}}}$. We also apply
 equation \eqref{eqn:DerD/2} to $D A_{2\nu^{\{l-1,l\}}}$ and apply
 equation~\eqref{eqn:DerDD/2} to $D^2 A_{2\nu^{\{l-1,l\}}}$.
 We then obtain the following equation
\begin{align*}
\Psi(\nu)
=&\sum_{j=1}^{l-2} (-1)^{j} A_{(2\nu_j,2\nu_{l-1})} \Psi(\nu^{\{j,l-1\}})
+\frac{1}{2}\sum_{i,j=1 \atop i \neq j}^{l-2} (-1)^{i+j+\delta_{j>i}} \Psi((\nu_i,\nu_j,\nu_{l-1},\nu_{l})) A_{2\nu^{\{i,j,l-1,l\}}} \nonumber\\
&+\frac{1}{4}\sum_{i,j=1 \atop i \neq j}^{l-2} (-1)^{i+j+\delta_{j>i}}
\Gamma((\nu_i,\nu_j,\nu_{l-1},\nu_{l}))
\cdot D A_{2\nu^{\{i,j,l-1,l\}}}
-\frac{1}{2} \Gamma(\nu^{\{l-1,l\}}) \cdot D A_{(2\nu_{l-1},2\nu_l)},
\end{align*}
where $\Gamma(\nu)$  is defined by equation \eqref{eqn:Gamma}.
By Lemma \ref{lem:Gamma}, $\Gamma((\nu_i,\nu_j,\nu_{l-1},\nu_{l}))=0$ and $\Gamma(\nu^{\{l-1,l\}})=0$.
Hence we have $\Psi(\nu)=0$ by induction.
The lemma is thus proved.
$\Box$

\vspace{6pt}
We are now ready to prove the main result of this subsection.

\noindent
{\bf Proof of Theorem \ref{thm:KW Schur Q}}: 
As explained after the proof of Proposition \ref{pro:Phi}, to prove Theorem \ref{thm:KW Schur Q}, it suffices to
show $\Phi(\mu)=0$ for all $\mu \in \widetilde{DP}$, where $\Phi(\mu)$ is defined by equation \eqref{eqn:Phi}.
We now prove $\Phi(\mu)=0$ by induction on $l=l(\mu)$, which is always even by assumption. 

First, it is straightforward to check $\Phi(\mu)=0$ using equations
\eqref{eqn:Ap2} and \eqref{eqn:Al<3} if $l=2$.

Assume $l \geq 4$.
As explained in the remark after equation \eqref{eqn:A rec}, we can
expand all terms appear on the right hand side of equation \eqref{eqn:Phi} 
using equation \eqref{eqn:A rec} with $n=l$. Then we obtain
\begin{align*} 
\Phi(\mu)
=\sum_{i=1}^{l-1}(-1)^{i+1} \Big[A_{(2\mu_i,2\mu_l)}\Phi(\mu^{\{i,l\}})
+A_{2\mu^{\{i,l\}}}\Phi((\mu_i,\mu_l))\Big]
+24\Psi(\mu),
\end{align*}
where $\Psi(\mu)$ is defined by equation \eqref{eqn:def Psi}. By Lemma \ref{lem:Psi}, $\Psi(\mu)=0$.
Hence by induction, we have $\Phi(\mu)=0$ since 
$l(\mu^{\{i,l\}})<l(\mu)$.
This completes the proof of Theorem \ref{thm:KW Schur Q}.
$\Box$

\section{Actions of W-type operators $\widetilde{P}^{(k)}_m$ on  Schur functions} \label{sec:pf1}

Throughout this section, we set $\rho=0$ and assume all operators are acting on $\mathbb{C}[\mathbf{t}]$. In this case, equations \eqref{eqn:commutation relation for HL:BB}
and \eqref{eqn:commutation relation for HL:B*B}  are reduced to
\begin{align}\label{eqn:BB=-BB}
B_aB_b=-B_{b-1}B_{a+1}
\end{align}
and
\begin{align}\label{eqn:BB*=-B*B+delta}
B_aB^*_b=-B^*_{b-1}B_{a-1}+\delta_{a,b}
\end{align}
for all $a, b \in \mathbb{Z}$. In particular, equation \eqref{eqn:BB=-BB} implies that
\begin{align} \label{eqn:Baa+1}
B_a B_{a+1} = 0
\end{align}
for all $a \in \mathbb{Z}$. We can use this formula to prove the following
\begin{lem} \label{lem:S-n1m}
For all $\la \in \mathbb{Z}^l$ and integers $m \geq 0$, $n \leq 0$,
\[ S_{(\la, n, 1^m)} = \delta_{m,-n} (-1)^m S_\la, \]
where $1^m=(1, \cdots, 1) \in \mathbb{Z}^m$.
\end{lem}
{\bf Proof}:
Let $\la=(\la_1, \cdots, \la_l)$. By definition of Schur functions,
\begin{equation} \label{eqn:SLn1}
S_{(\la, n, 1^m)} = B_{\la_1} \cdots B_{\la_l} B_{n} (B_1)^m \cdot 1.
\end{equation}

If $n \leq -m$, repeatedly applying equation \eqref{eqn:BB=-BB} to the right hand side of above equation $m$ times, we have
\begin{align*}
S_{(\la, n, 1^m)} =& (-1)^m B_{\la_1} \cdots B_{\la_l}  (B_0)^m B_{n+m} \cdot 1
= (-1)^m B_{\la_1} \cdots B_{\la_l}  (B_0)^m \cdot \delta_{m, -n}  \\
=& (-1)^m B_{\la_1} \cdots B_{\la_l}   \cdot \delta_{m, -n} = \delta_{m, -n}(-1)^m S_\la ,
\end{align*}
since $B_{k} \cdot 1= \delta_{k,0}$ for all $k \leq 0$.

If $0 \geq n > -m$, repeatedly applying equation \eqref{eqn:BB=-BB} to the right hand side of equation~\eqref{eqn:SLn1}
$(-n)$ times, we have
\begin{align*}
S_{(\la, n, 1^m)} =& (-1)^{-n} B_{\la_1} \cdots B_{\la_l}  (B_0)^{-n}  B_{0} B_{1} (B_1)^{m+n-1} \cdot 1
= 0 ,
\end{align*}
since $B_0 B_1 =0$.
The lemma is thus proved.
$\Box$

\subsection{Proof of Theorem \ref{thm:main1}}

Since $\rho=0$, Corollary \ref{cor:O  as BB} in this case has the following simple form,
\begin{align}\label{eqn:O=BBS}
\widetilde{P}^{(k)}_m
= -k \sum_{a,b\in\mathbb{Z}\atop a-b=m}  [b]_{k-1} \, \left( B_{a}^*B_{b}-\delta_{a,b} \delta_{b \geq 0} \right) +\delta_{m,0} \delta_{k,0},
\end{align}
for all integer $k \geq 0$ and $m \in \mathbb{Z}$. To prove Theorem \ref{thm:main1}, we need to compute
actions of $B_n$ and $B_n^*$ on Schur functions first.

% where
% \begin{align}\label{eqn:def:B* star B}
% B^*_{a}\star B_b
% =B_{a}^*B_{b}-\delta_{a,b} \delta_{b \geq 0}.
% \end{align}

By definition of Schur functions and equation \eqref{eqn:BB=-BB},
\begin{align}
 B_n \cdot S_{(\la,\mu)} =& B_n B_{\la_1} \cdots B_{\la_l} \cdot S_{\mu}
= (-1)^l  B_{\la_1-1} \cdots B_{\la_l-1} B_{n+l} \cdot S_\mu  \nonumber \\
=& (-1)^l S_{(\la-1^l, n+l, \mu)}  \label{eqn:BnS}
\end{align}
 for any $n \in \mathbb{Z}$, $\la=(\la_1, \cdots, \la_l) \in \mathbb{Z}^l$, and $\mu \in \mathbb{Z}^m$.

Action of $B_n^*$ on Schur functions is given by the following lemma.
\begin{lem}\label{lem:B*_n 1}
For any $n \in \mathbb{Z}$ and $\la \in \mathbb{Z}^l$,
\begin{align} \label{eqn:B*S}
B^*_n \cdot S_\la
=\delta_{n\leq -l} \cdot (-1)^n S_{(\la+1^l,1^{-n-l})}
+\sum_{i=1}^l \delta_{\la_i,n+i-1}(-1)^{i-1} S_{\la^{\{i\}}+(1^{i-1},0^{l-i})},
\end{align}
where  $0^k=(0, \cdots 0) \in \mathbb{Z}^k$, and
$\la^{\{i\}} \in \mathbb{Z}^{l-1}$ is the vector obtained from $\la$ by deleting its $i-$th component.
\end{lem}
{\bf Proof:}
We first prove a special case of this lemma where $l=0$, i.e.
\begin{align} \label{eqn:B*n1}
	B^*_{-n} \cdot 1 =\delta_{n\geq0} (-1)^n S_{(1^n)}
\end{align}
for any $n\in\mathbb{Z}$.

If $n \leq 0$, this equation holds trivially since
 $S_{\emptyset}=1$ and in this case
$B^*_{-n} \cdot 1=\delta_{n,0}$  (see, for example, Proposition 2.1 in \cite{JL}).
For $n > 0$, we prove this equation by induction on $n$.
Assume this equation holds for $n=k \geq 0$.
By equation \eqref{eqn:BB*=-B*B+delta}, we have
\begin{align*}
B^*_{-(k+1)} \cdot 1
=B^*_{-(k+1)} B_0 \cdot 1 = - B_1 B^*_{-k} \cdot 1=-B_1 \cdot (-1)^k S_{(1^k)} = (-1)^{k+1} S_{(1^{k+1})}.
\end{align*}
By induction, equation \eqref{eqn:B*n1} holds for all $n$.

We prove this lemma by induction on $l$. For $l>0$, assume the lemma holds for all $\la \in \mathbb{Z}^{l-1}$. Then
for any $\la \in \mathbb{Z}^{l}$, we write $\la=(\la_1, \la^{\{1\}})$. We also use $\la^{\{i,j\}}$ to denote
the vector obtained from $\la$ by deleting its $i$-th component and $j$-th component.

By equation \eqref{eqn:BB*=-B*B+delta}, we have
\begin{align*}
& B^*_{n} \cdot S_\la
=  B^*_{n} B_{\la_1} \cdot S_{\la^{\{1\}}} = (-B_{\la_1 + 1} B^*_{n+1} + \delta_{\la_1, n}) \cdot S_{\la^{\{1\}}} \\
=& B_{\la_1 + 1} \cdot \left\{ \delta_{n\leq -l} \cdot (-1)^{n} S_{(\la^{\{1\}}+1^{l-1},1^{-n-l})}
  - \sum_{i=1}^{l-1} \delta_{\la_{i+1},n+i}(-1)^{i-1} S_{\la^{\{1, i+1\}}+(1^{i-1},0^{l-1-i})} \right\}
  + \delta_{\la_1,n} S_{\la^{\{1\}}} \\
=& \delta_{n\leq -l} \cdot (-1)^{n} S_{(\la +1^{l},1^{-n-l})}
  - \sum_{i=1}^{l-1} \delta_{\la_{i+1},n+i}(-1)^{i-1} S_{\la^{\{i+1\}}+(1^{i},0^{l-1-i})} + \delta_{\la_1,n} S_{\la^{\{1\}}},
\end{align*}
which is equal to the right hand side of equation \eqref{eqn:B*S}.
Hence the lemma is proved.
$\Box$

\begin{rem}
If $\la$ is a partition, i.e. $\la_1 \geq \la_2 \geq \dots  \geq \la_l >0$, then all Schur functions appeared on the right hand side of
equation  \eqref{eqn:B*S} are  associated with partitions too.
\end{rem}

We are now ready to prove the second main result of this paper.

\vspace{6pt}
\noindent
{\bf Proof of Theorem \ref{thm:main1}}:
The theorem holds trivially for $k=0$. So we assume $k \geq 1$.

By equation \eqref{eqn:O=BBS} and equation \eqref{eqn:BB*=-B*B+delta}, we have
\begin{align*} %\label{eqn:O=BB*S}
\widetilde{P}^{(k)}_m
=& -k \sum_{a,b\in\mathbb{Z}\atop a-b=m}  [b]_{k-1} \, \left( - B_{b+1} B_{a+1}^*+\delta_{a,b}-\delta_{a,b} \delta_{b \geq 0} \right) \\
=& k \sum_{a,b\in\mathbb{Z}\atop a-b=m}  [b-1]_{k-1} \, \left( B_{b} B_{a}^*-\delta_{a,b} \delta_{b \leq 0} \right).
\end{align*}
for all integer $k \geq 1$ and $m \in \mathbb{Z}$. For any $\la=(\la_1, \cdots, \la_l) \in \mathbb{Z}^l$, by
Lemma \ref{lem:B*_n 1} and equation \eqref{eqn:BnS}, we have
\begin{align}
\widetilde{P}^{(k)}_m S_\la
=& k \sum_{a,b\in\mathbb{Z}\atop a-b=m}  [b-1]_{k-1} \, \bigg\{ -\delta_{a,b} \delta_{b \leq 0}  S_\la \nonumber \\
& \hspace{40pt}         + B_{b} \bigg( \delta_{a \leq -l} \cdot (-1)^a S_{(\la+1^l,1^{-a-l})}
            +\sum_{i=1}^l \delta_{\la_i,a+i-1}(-1)^{i-1} S_{\la^{\{i\}}+(1^{i-1},0^{l-i})} \bigg)
       \bigg\} \nonumber \\
 =& k \sum_{a,b\in\mathbb{Z}\atop a-b=m}  [b-1]_{k-1} \, \bigg\{ -\delta_{a,b} \delta_{b \leq 0} S_\la
           + \delta_{a \leq -l} \cdot (-1)^{a+l} S_{(\la, b+l,1^{-a-l})} \nonumber \\
& \hspace{100pt}             +\sum_{i=1}^l \delta_{\la_i,a+i-1}(-1)^{i+l} S_{(\la^{\{i\}}-(0^{i-1},1^{l-i}), b+l-1)}
        \bigg\}.   \label{eqn:PS1}
\end{align}
By equation \eqref{eqn:BB=-BB}, for any $1 \leq i \leq l$,
\begin{align*}
 S_{(\la^{\{i\}}-(0^{i-1},1^{l-i}), b+l-1)}
 =& B_{\la_1} \cdots B_{\la_{i-1}} B_{\la_{i+1}-1} \cdots B_{\la_l -1} B_{b+l-1} \cdot 1 \\
 =& (-1)^{l-i} B_{\la_1} \cdots B_{\la_{i-1}} B_{b+i-1}  B_{\la_{i+1}} \cdots B_{\la_l}  \cdot 1  \\
 =& (-1)^{l-i} S_{\la - (\la_i-b-i+1)\epsilon_i}.
\end{align*}
Hence, equation \eqref{eqn:PS1} is equivalent to
\begin{align}
 \widetilde{P}^{(k)}_m S_\la
=& \, k \, R_{\la, k ,m} + k \sum_{i=1}^l  [\la_i -i-m]_{k-1} S_{\la-m \epsilon_i},   \label{eqn:PS2}
\end{align}
where
\[ R_{\la, k ,m} :=\sum_{a,b\in\mathbb{Z}\atop a-b=m}  [b-1]_{k-1} \, \bigg\{ -\delta_{a,b} \delta_{b \leq 0} S_\la
           + \delta_{a \leq -l} \cdot (-1)^{a+l} S_{(\la, b+l,1^{-a-l})} \bigg\}. \]
Moreover, by Lemma \ref{lem:S-n1m}, if $b+l \leq 0$ and $a+l \leq 0$,
\begin{equation} \label{eqn:blal}
S_{(\la, b+l,1^{-a-l})} = \delta_{a,b} (-1)^{a+l} S_\la.
\end{equation}
If $m \neq 0$, then $\delta_{a,b}=0$ since $a-b=m$. Hence, by equation \eqref{eqn:blal},  we have
\begin{align}
R_{\la, k ,m}
=& \sum_{m-l<a \leq -l} [a-m-1]_{k-1} \, (-1)^{a+l}  S_{(\la, a-m+l,1^{-a-l})}  \nonumber \\
=& \delta_{m<0} \sum_{n=1}^{-m} (-1)^{n+m} [n-l-1]_{k-1} \, \cdot  S_{(\la, n,1^{-n-m})},  \label{eqn:Rm+-}
\end{align}
where we have set $n = a+l-m$ in the second equality.

If $m=0$,  by equation \eqref{eqn:blal}, we have
\begin{align}
R_{\la, k ,0}
=& \sum_{a \in \mathbb{Z}}  [a-1]_{k-1} \, \bigg\{ - \delta_{a \leq 0} S_\la
           + \delta_{a \leq -l} \cdot (-1)^{a+l} S_{(\la, a+l,1^{-a-l})} \bigg\} \nonumber \\
=& - \left( \sum_{-l < a \leq 0 }    [a-1]_{k-1} \right) \,  S_\la.  \nonumber
\end{align}
Moreover, replacing $a$ by $-n$, we have
\begin{align*}
\sum_{-l < a \leq 0 } [a-1]_{k-1}
=& (-1)^{k-1} \sum_{n=0}^{l-1} \frac{(n+k-1)!}{n!} = (-1)^{k-1} (k-1)! \sum_{n=0}^{l-1} \binom{n+k-1}{k-1} \\
=& (-1)^{k-1} (k-1)! \binom{l+k-1}{k} = - \frac{1}{k} [-l]_k,
\end{align*}
where in the third equality, we have used the well known combinatorial identity
\[ \sum_{m=0}^n \binom{m}{k} = \binom{n+1}{k+1} \]
which is a consequence of Chu-Vandermonde identity. Hence we have
\begin{align} \label{eqn:Rm0}
R_{\la, k ,0}
=& \frac{1}{k} [-l]_k \,  S_\la.
\end{align}

Combining equations \eqref{eqn:PS2}, \eqref{eqn:Rm+-}, \eqref{eqn:Rm0}, we obtain the desired formula.
This completes the proof of Theorem \ref{thm:main1}.
$\Box$

\subsection{Special cases of Theorem \ref{thm:main1}}

Since $\rho=0$, the operator $J(z)$ has the form
\begin{align*}
J(z) =\sum_{n\in\mathbb{Z}_{+}} nt_n \, z^{n-1} +\sum_{n\in\mathbb{Z}_{+}} \frac{\partial}{\partial t_n} \, z^{-n-1}.
\end{align*}
\begin{exa}
	Consider the $k=1$ case. Since $P^{(1)}(z)=J(z)$, Theorem \ref{thm:main1} in this case is equivalent to
	\begin{align*}
	\frac{\partial}{\partial t_m} \cdot  S_\la
	=\sum_{i=1}^l S_{\la-m\epsilon_i}, \,\,\,\,\,\, {\rm and} \,\,\,\,\,\,
	mt_m \cdot  S_\la=\sum_{i=1}^l S_{\la+m\epsilon_i}+\sum_{n=1}^{m} (-1)^{m-n} S_{(\la,n,1^{m-n})}
	\end{align*}
    for $m>0$. These formulas are well-known. The formula for $t_m \cdot  S_\la$ is related to the Murnaghan-Nakayama
    rule for calculating characters of irreducible representations of the symmetric group
    (see, for example, Chapter I in \cite{Mac}).
\end{exa}

\begin{exa}
	Consider the $k=2$ case. Let $L_m$ be the coefficient of $z^{-m-2}$ in $\frac{1}{2}:J(z)^2:$. Then
	\begin{align*}
	L_m:=&\frac{1}{2}\widetilde{P}^{(2)}_m - \frac{-m-1}{2}\widetilde{P}^{(1)}_m.
	\end{align*}
	More explicitly,
	\begin{align*}
	L_m=\sum_{a>\max\{0,-m\}} at_a\frac{\partial}{\partial t_{a+m}}
	+\frac{\delta_{m>0}}{2}\sum_{k=1}^{m-1} \frac{\partial^2}{\partial t_k \partial t_{m-k}}
	+ \frac{\delta_{m<0}}{2}\sum_{k=1}^{-m-1} k(-m-k)t_k t_{-m-k}.
	\end{align*}
    These operators generate a Virasoro algebra.
	By Theorem \ref{thm:main1},
	\begin{align*}
	L_m \cdot  S_\la =&\sum_{i=1}^l \big(\la_i-i+\frac{1-m}{2}\big) S_{\la-m\epsilon_i}\\
	&+\delta_{m<0}\sum_{n=1}^{-m} (-1)^{m-n}\big(n-l+\frac{m-1}{2}\big) S_{(\la,n,1^{-m-n})}
	+\delta_{m,0}\frac{l^2}{2} S_\la.
	\end{align*}
	A formula for the action of Virasoro operators on Schur functions was given in Theorem 5.1 in \cite{AM}. However, coefficients
in that formula are very complicated and are written as summations over special sets of partitions. A simpler formula, which agrees with the formula here, was given in \cite{LY3}. Some parts of these formulas are used to study Gaussian Hermitian matrix models in \cite{MMMR}.
\end{exa}

\begin{exa}
 Consider the $k=3$ case.
Let $W_m$ the the coefficient of $z^{-m-3}$ in $\frac{1}{3}:J(z)^3:$. Then
\begin{align}\label{eqn:W_m}
W_m:=&\frac{1}{3}\widetilde{P}^{(3)}_m-\frac{-m-2}{2}\widetilde{P}^{(2)}_m+\frac{(m+1)(m+2)}{6}\widetilde{P}^{(1)}_m.
\end{align}
By Theorem \ref{thm:main1},
\begin{align*}
W_m S_\la
=&\sum_{i=1}^l \left( \frac{m^2}{6}+im-\la_im-\frac{m}{2}+i^2-2i\la_i-i+\la_i^2+\la_i+\frac{1}{3} \right) S_{\la-m\epsilon_i}\\
&+\delta_{m<0}\sum_{n=1}^{-m} (-1)^{m-n} \left( \frac{m^2}{6}+mn-lm-\frac{m}{2}+n^2-2ln-n+l^2+l+\frac{1}{3} \right) S_{(\la,n,1^{-m-n})}\\
&+\delta_{m,0}\left( -\frac{l^3}{3} \right) S_\la.
\end{align*}
% \begin{pro}
% For the special $m=0$ case, we write it more explicitly as
% \begin{align*}
% W^A_0\cdot S_\la= \sum_{i=1}^l (\la_i^2+\la_i-2i\la_i) S_\la,
% \end{align*}
% where
The action of operator
\begin{align*}
W_0=\sum_{a,b\in\mathbb{Z}_{>0}}
\bigg(abt_at_b\frac{\partial}{\partial t_{a+b}}
+(a+b)t_{a+b}\frac{\partial^2}{\partial t_{a} \partial t_b}\bigg)
\end{align*}
on Schur functions was studied in \cite{G}. Two proofs of the formula for $W_0 S_\la$
were given in \cite[Proposition 2]{FW} and \cite[Proposition 2.2]{Zhou}.
Formula for $W_0 S_\la$ was used in \cite{KL} to express the generating function of Hurwitz numbers in terms of Schur
functions, where $W_0$ was also called the cut-and-join operator.
Modified  formulas for $W_{-1} S_\la$ and $W_{-2} S_\la$  were used in \cite{MMMR} and \cite{WZZZ} to study the Gaussian Hermitian matrix model.
\end{exa}

\section{Conflict of interest and data availability statement}
On behalf of all authors, the corresponding author states that there is no conflict of interest, and
no datasets were generated or analysed during the current study.

%%%%%%%%%%%%%%%%%%%%%%%%%%%%%%%%%%%%%%%%%%%%%%%%%%%

\vspace{30pt} \noindent
Xiaobo Liu \\
School of Mathematical Sciences \& \\
Beijing International Center for Mathematical Research, \\
Peking University, Beijing, China. \\
Email: {\it xbliu@math.pku.edu.cn}
\ \\ \ \\
Chenglang Yang \\
Hua Loo-Keng Center for Mathematical Sciences \& \\
Academy of Mathematics and Systems Science, \\
Chinese Academy of Sciences, Beijing, China. \\
Email: {\it yangcl@amss.ac.cn}

\end{document}